\documentclass[a4paper,fleqn]{cas-dc}

\usepackage[numbers]{natbib}
\usepackage{caption}
\usepackage{adjustbox}
\usepackage{wrapfig}
\usepackage{graphicx}
\usepackage{afterpage}
\usepackage{amsmath,amssymb,amsfonts,utfsym,pifont,tikz,xparse}

\def\tsc#1{\csdef{#1}{\textsc{\lowercase{#1}}\xspace}}
\tsc{WGM}
\tsc{QE}
\tsc{EP}
\tsc{PMS}
\tsc{BEC}
\tsc{DE}

\begin{document}\sloppy
\let\WriteBookmarks\relax
\def\floatpagepagefraction{1}
\def\textpagefraction{.001}
\shorttitle{}
\shortauthors{Yilei Zhang et~al.}
\title [mode = title]{EFACT: an External Function Auto-Completion Tool to Strengthen Static Binary Lifting}

\author{Yilei Zhang}
\author{Haoyu Liao}
\author{Zekun Wang}
\author{Bo Huang}[type=editor,orcid=0000-0001-5126-7192]
\cormark[1]
\ead{bhuang@dase.ecnu.edu.cn}
\author{Jianmei Guo}
\address{School of Data Science and Engineering, East China Normal
University, Shanghai 200062, China}

\begin{abstract}
Static binary lifting is essential in binary rewriting frameworks. Existing tools overlook the impact of External Function Completion (EXFC) in static binary lifting. EXFC recovers the declarations of External Functions (EXFs, functions defined in standard shared libraries) using only the function symbols available. Incorrect EXFC can misinterpret the source binary, or cause memory overflows in static binary translation, which eventually results in program crashes. Notably, existing tools struggle to recover the declarations of mangled EXFs originating from binaries compiled from C++.  Moreover, they require time-consuming manual processing to support new libraries.
 
This paper presents EFACT, an External Function Auto-Completion Tool for static binary lifting. Our EXF recovery algorithm better recovers the declarations of mangled EXFs, particularly addressing the template specialization mechanism in C++. EFACT is designed as a lightweight plugin to strengthen other static binary rewriting frameworks in EXFC. Our evaluation shows that EFACT outperforms RetDec and McSema in mangled EXF recovery by 96.4\% and 97.3\% on SPECrate 2017.
 
Furthermore, we delve deeper into static binary translation and address several cross-ISA EXFC problems. When integrated with McSema, EFACT correctly translates 36.7\% more benchmarks from x86-64 to x86-64 and 93.6\% more from x86-64 to AArch64 than McSema alone on EEMBC.
\end{abstract}

\begin{keywords}
external function completion \sep binary lifting \sep static binary translation \sep reverse engineering
\end{keywords}

\maketitle

\section{Introduction}\label{introduciton}

Binary lifting~\cite{DBLP:journals/csur/WenzlMUW19} is a critical process in various domains, such as binary analysis~\cite{DBLP:conf/cav/BrumleyJAS11,verbeek2022formally,DBLP:journals/jss/SaievaK22}, binary translation~\cite{ GUAN2012305,FU2019173,DBLP:journals/jss/WuDFZWZ22}, binary optimization~\cite{bala2000dynamo,dyninst,al2019binary}, and binary hardening~\cite{di2017rev}. As machine code is humanly unreadable, it is essential to uplift it into a higher-level representation. Generally, there are three categories of binary lifting: binary-to-assembly (ASM), binary-to-intermediate representation (IR), and binary-to-high-level languages, e.g. C/C++.

\begin{figure*}
\centering
\centerline{\includegraphics[height=4.8cm]{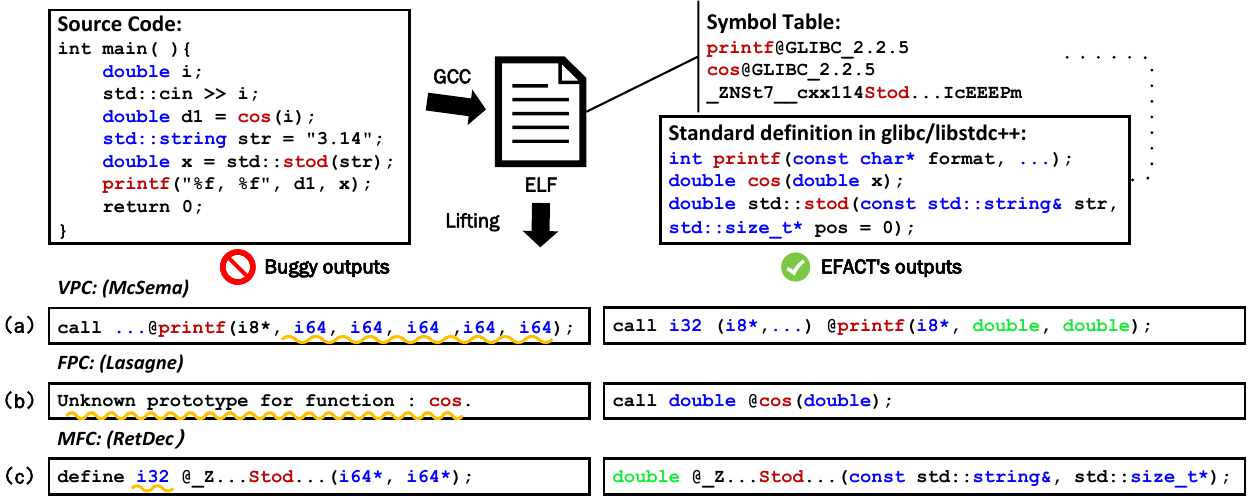}}

\caption{Motivating examples of EXFC (source code is unknown in actual scenarios, we put it here to better explain the example).}
\label{motivatingExample}
\end{figure*}

Static analysis of binaries has long been considered a challenging task, leading to the development of dynamic binary frameworks ~\cite{anand2013compiler}. However, the high cost associated with a dynamic environment makes it unsuitable for certain scenarios. In the framework of static binary lifting, the External Function Completion (EXFC) problem is inevitable when selecting IR and high-level languages as lifting outputs. An External Function (EXF) is a function defined in an external shared library, such as \textit{glibc}. Fig.~\ref{motivatingExample} shows some examples of EXFC, and let's take ``\textit{cos}'' in Fig.~\ref{motivatingExample} as an example. When the function named ``\textit{cos}'' appears in an Executable and Linkable Format (ELF) file's (dynamically linked) symbol table, the full declaration, ``\textit{double cos (double x)}'', should be returned. For binary analysis, a failure in EXFC can hinder the understanding of the binary. Meanwhile, for static binary translation, it may trigger precision errors or memory overflows, which eventually lead to severe consequences, such as program crashes.
 
Current static binary lifting frameworks exhibit certain deficiencies in addressing the EXFC problem, which can be broadly categorized into the following two aspects:

\newdefinition{definition}{Deficiency}
\begin{definition}
\textbf{Incorrect recovery of function parameters and return type.} Fig.~\ref{motivatingExample}, parts (a) and (c), evaluate the binary lifting capabilities of two state-of-the-art static binary lifting tools~\cite{DBLP:conf/sp/LiuYWB22}, McSema~\cite{McSema} and RetDec~\cite{kvroustek2017retdec} (based on Capstone~\cite{capstone}). We use two EXFs for evaluation, \textit{stod} and \textit{printf}, which correspond to mangled function completion (MFC) and variable-parameter function completion (VPC), respectively. Initially, we compile the code using the GCC compiler with the dynamic linking option enabled. Subsequently, we submit the generated ELF file to McSema and RetDec. As illustrated in the bottom left corner of Fig.~\ref{motivatingExample}, McSema’s lifting result of \textit{printf} contains five \textit{i64} type parameters, whereas our source code specifies two \textit{double} type. RetDec's output indicates an \textit{i32} as the return type of \textit{stod}, while the correct return type is \textit{double}. McSema and RetDec are both used as binary analysis tools, but the lifting results mislead the understanding of the source binary. This straightforward experiment illustrates that existing tools still face challenges in recovering the function declarations of variable-parameter functions and mangled functions.

\end{definition}
\begin{definition}
\textbf{Difficulty in integrating a new external library or migrating to another framework.} In part (b) of Fig.~\ref{motivatingExample}, we evaluate the EXFC capabilities of a static binary translator, Lasagne~\cite{rocha2022Lasagne}, which is based on Microsoft’s llvm-mctoll~\cite{10.1145/3316482.3326354}. For this test, we use the \textit{cos} function, representing a simple example of fixed-parameter function completion (FPC). While Lasagne fails to recover, it's important to note that recovering \textit{cos} alone is not a difficult task, but achieving comprehensive EXFC entails considerable effort. Lasagne mainly focuses on addressing the strong/weak memory model issue, which indeed is an important research problem in binary translation. As Lasagne researchers haven't taken EXFC as part of their research, Lasagne's shortcomings in EXFC significantly diminish its overall usability. Besides, our investigation of existing open-source static binary lifting tools reveals that the source code portion responsible for handling EXFC is tightly coupled with other components. Therefore, users must possess a deep understanding of the project structure to support a new library (such as OpenSSL~\cite{OpenSSL}) or to migrate to another binary lifting framework that currently lacks EXFC support. Although McSema and Lasagne offer an interface to manually supply the declarations of EXFs, the complexity and vastness of external libraries make this task formidable. A tool capable of auto-generating declarations can significantly enhance the accuracy and efficiency of the current static lifting process.
\end{definition}
 
In response to the aforementioned limitations, we propose EFACT, an External Function Auto-Completion Tool designed for static binary lifting.
 
\textbf{To address deficiency 1}, first, we design an MFC algorithm to more accurately recover the return type of mangled EXFs in C++. While the standard LLVM's Demangling API\footnote{https://llvm.org/doxygen/Demangle\_8h\_source.html} is a frequently used API for obtaining detailed function declarations of mangled EXFs, its output, as illustrated in Fig.~\ref{C++AutoComple} (a), does not consistently guarantee the accurate identification of the function return type~\cite{itanium-Demangle}; only 13.5\% of mangled EXFs in SPEC CPU® 2017 can get a return type through this API. Furthermore, as depicted in Fig.~\ref{motivatingExample}, current static binary lifting tools incorrectly treat all unknown cases as \textit{int} (in C/C++) or \textit{i32}/\textit{i64} (in LLVM IR~\cite{DBLP:conf/cgo/LattnerA04}), offering no deeper recovery mechanism. Our MFC algorithm, not only recovers the implicit \textit{this} parameter but also provides a mechanism to deal with the template specialization mechanism in C++. 
 
Second, we develop a \textit{Dict} auto-generator that extracts complete function declarations from external libraries and catalogs them in a dictionary (\textit{Dict}). This not only supplements the information missing from the LLVM’s Demangling API but also aids in resolving VPC and FPC issues more effectively. The lower right corner of Fig.~\ref{motivatingExample} illustrates an example of EFACT's outputs, and in the motivating example, our tool outperforms three others. 
 
Third, in the process of binary lifting, careful consideration should be taken for the input binary file, including instruction set architecture (ISA), processor architecture, operating system (OS), and library version. Existing tools overlook the impact of these dimensions in their EXFC component. Based on our \textit{Dict} auto-generator, we construct a Library Database, which is organized layer by layer, from ISA (x86 or ARM) to \textit{OS\_Type} (CentOS or Ubuntu), and finally \textit{OS\_Version} (Ubuntu18.04 or 20.04), ensuring extensive coverage across multiple dimensions. By leveraging backward compatibility, we ingeniously cover the processor architecture and library version dimensions.
 
\textbf{To address deficiency 2}, EFACT necessitates only an ELF file as input and is capable of generating outputs in both LLVM IR and C/C++ program formats, making it an ideal plugin for binary-to-IR and binary-to-high-level language lifting frameworks. This helps enhance analysis and migration capabilities.
 
To make our solution more general, our tool is designed to be easily extended to embrace binaries compiled from other programming languages or binaries containing EXFs defined in other libraries. Our auto-completion workflow maximizes the utilization of the standard compiler toolchain, compilers of most programming languages generate an object file containing information akin to a symbol table, and our approach is effective as long as the ``symbol table'' is available.
 
In this paper, we select binaries compiled from C/C++ source programs along with the \textit{glibc}/\textit{libstdc++} as our framework's initial subjects, also we use binaries compiled from Fortran and Rust to test the extensibility. We choose SPEC CPU® 2017~\cite{SPEC2017} as our evaluation benchmark to evaluate whether EFACT can cover 100\% of the EXFs from \textit{glibc}, \textit{libstdc++} and \textit{libgfortran} in it. We delve further into the static binary translation domain and evaluate whether a binary translator, augmented with EFACT, can enhance the accuracy of static binary translation. This also demonstrates the usability of our tool. Overall, our paper makes the following key contributions:
\begin{itemize}
\item We recognize that existing static binary lifting tools overlook the impact of failure in EXFC. To better address this problem, we introduce EFACT, an External Function Auto-Completion Tool that generates complete function declarations of EXFs presented in a target ELF file. Different from other similar tools, our tool can recover the return type and the implicit \textit{this} parameter of mangled EXFs based on our MFC algorithm and automatically generated dictionaries (\textit{Dicts}) which cover popularly used external function declarations, showcasing greater robustness compared to the state-of-the-art tools, such as RetDec and McSema~\cite{DBLP:conf/sp/LiuYWB22}. Additionally, we delve further into the static binary translation domain, addressing several cross-ISA EXFC problems.
\item EFACT is publicly available\footnote{https://github.com/Stephen-lei/EFACT} and can be integrated as a plugin in other binary rewriting frameworks. We evaluate its usability with McSema and Lasagne. EFACT is easily extensible to support binaries compiled from other languages (Fortran and Rust) and binaries containing EXFs defined in other libraries (OpenSSL). Furthermore, leveraging our innovative \textit{Dict} auto-generator, we establish a comprehensive Library Database that capitalizes on backward compatibility to encompass multiple dimensions, including ISA, processor architecture, OS, and library version.
\item For SPEC CPU® 2017, EFACT achieves 100\% coverage for lifting the EXFs source from \textit{glibc}, \textit{libstdc++} and \textit{libgfortran}. Notably, in the context of MFC, EFACT accurately recovers 96.4\% more function return types than RetDec and 97.3\% more than McSema on SPECrate 2017. Furthermore, we integrate EFACT into McSema and introduce EFACT\_MC,  our static binary translator. This combination facilitates a 36.7\% increase in successfully translating binaries from x86-64 to x86-64 and a 93.6\% increase from x86-64 to AArch64 for benchmarks on EEMBC~\cite{eembc2} than the original McSema.
\item We carry out a comprehensive analysis of SPEC CPU® 2017 with a focus on EXF generation. Our analysis demonstrates that factors such as ISA, optimization options, and processor architecture can impact the number of generated EXFs.
\end{itemize}

\section{Background}\label{background}
To date, researchers have not reached a unified consensus on the output type of binary lifting, and in this paper, the terminology of EXF is different from that of the ``\textit{extern}'' keyword in C/C++. In this section, we first present our understanding of binary lifting. Subsequently, we provide a comprehensive explanation of what is EXFC, and why FPC, VPC, and MFC are challenging tasks, especially in the domain of static binary analysis and static binary translation. 
\subsection{Binary lifting}
\begin{figure}
\centerline{\includegraphics[height=3.5cm]{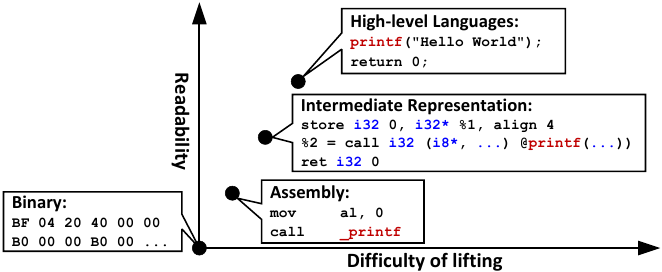}}
\caption{Layers of binary lifting.}
\label{liftingLayout}
\end{figure}

As illustrated in Fig.~\ref{liftingLayout}, the process of layer-by-layer abstraction enhances the readability of the extracted information, albeit at the expense of increased abstraction complexity. While numerous researchers refer to the abstraction from binary to
IR (primarily LLVM IR) as binary lifting, we align with Baldoni’s perspective~\cite{baldoni2018survey} that lifting is the process
of transforming a more detailed representation into a more
abstract one. Generally, we divide binary lifting into three categories: binary-to-ASM, binary-to-IR, and binary-to-high-level languages.

\subsection{External function completion and three challenging tasks}\label{THREE CHALLANGE}
Function completion involves recovering a detailed declaration of a function, including its return type and parameters. EXFs denote functions sourced from standard shared libraries, such as \textit{glibc}/\textit{libstdc++}. While source code contains information on parameters and return type, binary code only includes the function name in the symbol table. Recovering these function details during lifting not only enhances our understanding of the source binary but also is imperative for an IR-based or high-level language-based binary rewriting framework.
 
The main challenges in EXFC stem from three aspects:
 
\textbf{FPC (fixed-parameter function completion):} Functions such as \textit{sin} and \textit{cos} feature fixed declarations within the \textit{glibc}, allowing us to obtain the required information directly. 
For FPC, the central challenge lies in developing a method that can swiftly and efficiently encompass the entire standard library.

\begin{figure}
\centerline{\includegraphics[height=2cm]{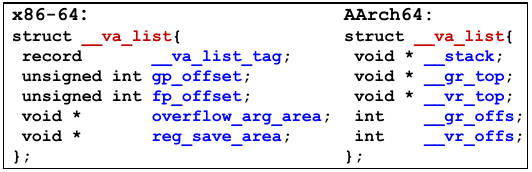}}
\caption{An abstraction of the detailed implementation of \textit{va\_list} on x86-64 and AArch64 (with GCC 9.4, Ubuntu 20.04).}
\label{DifferenceValist}
\end{figure}

\textbf{VPC (variable-parameter function completion):} A prevalent example of a variable-parameter function is \textit{printf}, which possesses the following standard declaration:
 
\textit{int printf (const char *\_\_restrict \_\_fmt, ...);}\\The ``...'' part is related to the actual binary context. Binary lifting tools need to examine the binary context before the invocation of \textit{printf} to identify real argument types and numbers. Notably, in cross-ISA static binary translation, some detailed implementation of variable-parameter functions varies across different ISAs. For example, \textit{va\_list} is a type used in C/C++  to manage variable argument lists. Although its initialization method is identical on x86 and ARM platforms, as depicted in the declaration:
 
\textit{void va\_start ( std::va\_list ap, format);}\\The concrete implementation within the data structure differs considerably. Fig.~\ref{DifferenceValist} shows an abstraction of how \textit{va\_list} is defined on x86\-64 and AArch64 (with GCC 9.4, Ubuntu 20.04). This variance in implementation details leads to discrepancies in how \textit{va\_list} appears on the stack. Failure to synchronize data on the stack can result in corruption in the translated binary.

\begin{figure*}
\centerline{\includegraphics[height=5cm]{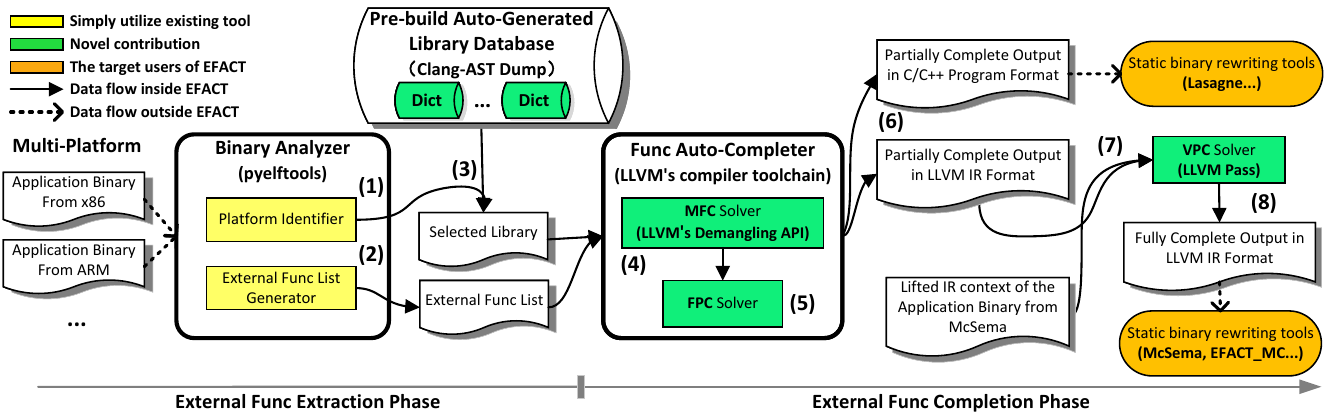}}
\caption{An overview of EFACT's auto-completion workflow (Func: Function).}
\label{EFACTFramework}
\end{figure*}
 
\textbf{MFC (mangled function completion):} Name mangling is a distinctive phenomenon in C++. We can find functions like ``\textit{\_ZNSolsEPFRSoS\_E}'' in the symbol table. Without demangling the function, identifying a corresponding function name in the standard library becomes virtually impossible. While the LLVM's compiler toolchain offers a Demangling API, it does not guarantee the discernment of the function's return type - only 13.5\% of mangled EXFs in SPEC CPU® 2017 can get a return type through this API, leading binary lifting tools inaccurately address all unidentified cases as either \textit{int} (in C/C++) or \textit{i32}/\textit{i64} (in LLVM IR), a flaw that can manifest in translation errors, such as casting a \textit{double} return type to an \textit{int}. This mismatching can potentially cause the translated program to crash during execution. The main reason stopping them from going deeper is the template specialization mechanism in C++. For instance, a preliminary analysis might reveal \textit{\_CharT} as a mangled EXF's return type from the standard library; however, determining whether to substitute it with \textit{char16\_t} or \textit{char32\_t} necessitates further investment. Overall, recovering the lost return type information, particularly for those involving template specialization, presents a considerable challenge.
 
\section{Framework overview }\label{frameworkovw}

\newcommand*\circled[1]{\tikz[baseline=(char.base)]{
\node[shape=circle, draw, inner sep=0.9pt, minimum size=1.5ex, text height=1.25ex, text depth=0.1ex] (char) {#1};}}
In this section, we will outline the complete structure of EFACT and explain how its different parts are connected. EFACT works on the x86-64 (with Ubuntu 20.04) platform and can handle ELF files from different platforms, like x86-64 (with CentOS 8) or AArch64 (with Ubuntu 20.04), which is practical for real-world tasks. EFACT automatically performs a static analysis of the ELF file and provides the recovered function declarations of the EXFs within the ELF file. Fig.~\ref{EFACTFramework} shows the overall process of EFACT, which is divided into two main parts: the External Function Extraction Phase and the External Function Completion Phase. We will go through each step of this process in detail.

\subsection{External function extraction phase}
\textbf{Step (1):} (Section \ref{platform identifier}) EFACT employs \textit{pyelftools}~\cite{pyelftools} to get the specific platform details of the source ELF file. These details include the ISA, processor architecture, OS, and information about the toolchain used to produce the binary. These details are crucial for selecting the appropriate external library versions from the Library Database. The reason why EFACT contains the step of choosing a specific version of the external library will be introduced in step (3).
 
\textbf{Step (2):} (Section \ref{EXF list generator}) EFACT employs \textit{pyelftools} to get the symbol table of the source ELF file. Then EFACT traverses the symbol table, extracts function symbols originating from the external standard library, and generates an EXF list. Moreover, EFACT identifies the programming language employed in the development of the source program. This step is crucial for determining the necessity of activating the MFC Solver (referenced in step (4), applicable exclusively for C++).
 
\textbf{Step (3):} (Section \ref{librarydatabases}) Based on the platform information obtained in step (1), EFACT selects the appropriate library version and \textit{Dict} from the Library Database. 
This Library Database is pre-built as preparation for practical use. The \textit{Dict} is generated automatically, and EFACT uses a script based on Clang-AST Dump\footnote{https://clang.llvm.org/docs/IntroductionToTheClangAST.html} to implement it. The \textit{Dict} plays an important role in the subsequent process of function completion. Another reason why EFACT contains such a Library Database is because EFACT is designed to maximize the use of the standard compiler toolchain. While the definition of a function remains consistent across different ISAs, despite varying implementation details, using a library built for x86 to recover an ELF file from an ARM platform can lead to numerous errors. Thus, the Library Database significantly enhances the flexibility and extensibility of our tool by accommodating diverse platforms.
 
\subsection{External function completion phase}\label{EFCPfRAME}
 
\textbf{Step (4):} (Section \ref{MFC Solver}) After obtaining the EXF list, EFACT deploys various strategies for function completion, taking into account the programming language used to develop the source program. As introduced in Section \ref{THREE CHALLANGE}, three primary challenges in EXFC are FPC, MFC, and VPC. At this stage, EFACT first activates the MFC Solver if C++ is identified as the programming language of the source program. Contrasting with other static binary lifting tools that depend exclusively on the output from LLVM’s Demangling API, EFACT additionally utilizes an MFC algorithm on this output. This strategy enhances EFACT's ability to more accurately recover function declarations, particularly the return types, of mangled functions.

\textbf{Step (5):} (Section \ref{FPC SOLVER}) Apart from the challenges associated with name mangling in C++, EFACT's methodology for handling C++ is consistent with the approaches used for other programming languages, such as C, Fortran, and Rust. EFACT employs the FPC Solver to recover the function declarations of functions with fixed parameters.

\textbf{Step (6):} EFACT generates two kinds of outputs: C/C++ program code and LLVM IR. At this stage, both kinds of outputs produced by EFACT are partially complete, as they do not address the VPC challenge due to the absence of application context. Currently, the partially complete output in C/C++ program format is not designed for further exploration into VPC, as EFACT's primary focus is on binary-to-IR level static binary lifting. However, the partially complete output in C/C++ program format is already usable as input to static binary rewriting tools, like Lasagne. The partially complete output in LLVM IR format, on the other hand, will continue to undergo further recovery processes.

\textbf{Step (7):} (Section \ref{VPC SOLVER}) Addressing the VPC challenge requires more than just the function declarations from the standard library. Additional information from the actual binary context is essential. EFACT's VPC Solver leverages the lifted IR context of the application binary, obtained from McSema, as a supplement to the real binary context. EFACT then employs several LLVM Passes (written by ourselves) to recover the function declarations of variable-parameter functions.
 
\textbf{Step (8):} To date, EFACT successfully generates a fully complete output in LLVM IR format. This output resolves the challenges of VPC, MFC, and FPC, and can be used for binary-to-IR level static binary rewriting tools, such as McSema and EFACT\_MC.
 
Our tool is inspired by McSema's \textit{generate\_abi.py} and we have made EFACT publicly accessible. Our enhancements to McSema's \textit{generate\_abi.py} can be summarized as follows:
\begin{itemize}
\item The old ABI completion tool in McSema has remained unmaintained for an extended period, rendering it non-functional. We rewrite this tool, making it functional and expanding its coverage beyond just the ISA perspective to encompass aspects of the OS, processor architecture, and library version. 
\item The old ABI completion tool in McSema requires users to supply the header files potentially utilized by the ELF file, followed by an exhaustive extraction of these files. The comprehensive extraction results are then integrated into the lifted IR, aiding McSema in more effectively lifting the source ELF file. However, it is often unclear which header files might have been used. Additionally, this extensive merging process resulted in a large volume of code in the lifted IR, despite the actual usage of only a few function declarations. In contrast, EFACT eliminates the requirement for users to provide a specific set of header files. We have developed a mechanism for extracting EXFs and automatically matching them with the corresponding function declarations in the standard library. EFACT specifically extracts the EXFs presented in the source ELF file, thereby avoiding the recovery of redundant functions. This approach is more aligned with practical scenarios and enhances usability.
\item Our enhanced tool improves the auto-completion of binaries compiled from C++ programs by effectively resolving the challenges associated with C++ name mangling. Moreover, in the domain of reverse engineering, we frequently encounter ELF files that are either quite dated (developed in languages like Fortran) or extremely recent (developed in languages like Rust). We believe that a robust EXFC framework should not be limited to just C/C++—as extensibility is crucial. In comparison to the older ABI completion tool in McSema, EFACT takes a more comprehensive stance regarding language coverage. The auto-completion capabilities of EFACT have been validated with binaries compiled from Fortran and Rust source programs, which have never been tried by other researchers.
\item In addition to generating LLVM IR output, our tool also generates C/C++ program output, which can be utilized by a  wider range of binary rewriting tools.
\end{itemize}

\section{External function extraction phase}\label{EFEP}
In this section, we describe how we determine the original platform of the target ELF file and build the Library Database. 
\subsection{Platform identifier}\label{platform identifier}
Our tool utilizes \textit{pyelftools} to identify the ISA, processor architecture, and OS of the target ELF file. Table \ref{tab1} displays the output of the motivating example in Section \ref{introduciton}. The \textit{Machine} and \textit{Class} items in the ELF Header carry information about ISA and processor architecture. From the \textit{.comment} section, we can obtain detailed information about the OS and compiler toolchain. This information assists us in locating the appropriate platform library.
\subsection{External function list generator}\label{EXF list generator}
Our tool traverses the symbol table of the target ELF file. There are two parts in the symbol table: the \textit{.symtab} section contains all the symbols, while the \textit{.dynsym} section contains all the dynamically linked ones. EFACT traverses both sections to cover the completable functions as much as possible. For stripped ELF, EFACT can still get the \textit{.dynsym} section to go on our completion.
 
\begin{table}
\centering
\caption{An example of pyelftools' output.}
\renewcommand{\arraystretch}{1.2}
\resizebox{0.37\textwidth}{!}{
\begin{tabular}{lllll}

\hline
\multicolumn{1}{l}{\textbf{Item}}&\multicolumn{1}{c}{\textbf{Result}} \\
\cline{1-2}
\multicolumn{1}{l}{Class} &\multicolumn{1}{l}{ELF64} \\
\multicolumn{1}{l}{Machine}&\multicolumn{1}{l}{Advanced Micro Devices x86-64} \\
\multicolumn{1}{l}{.comment}&\multicolumn{1}{l}{GCC: (Ubuntu 9.4.0-1ubuntu1~20.04.1) 9.4.0} \\
\multicolumn{1}{l}{.dynsym}&\multicolumn{1}{l}{cos@GLIBC\_2.2.5, \_ZSt3cin@GLIBCXX\_3.4}\\
\multicolumn{1}{l}{.symtab}&\multicolumn{1}{l}{\_ZNSt7\_\_cxx114stod...IcEEEPm}\\

\hline

\end{tabular}
\label{tab1}

}
\end{table}

\subsection{Library Database}\label{librarydatabases}
\begin{figure}
\centerline{\includegraphics[height=6.1 cm]{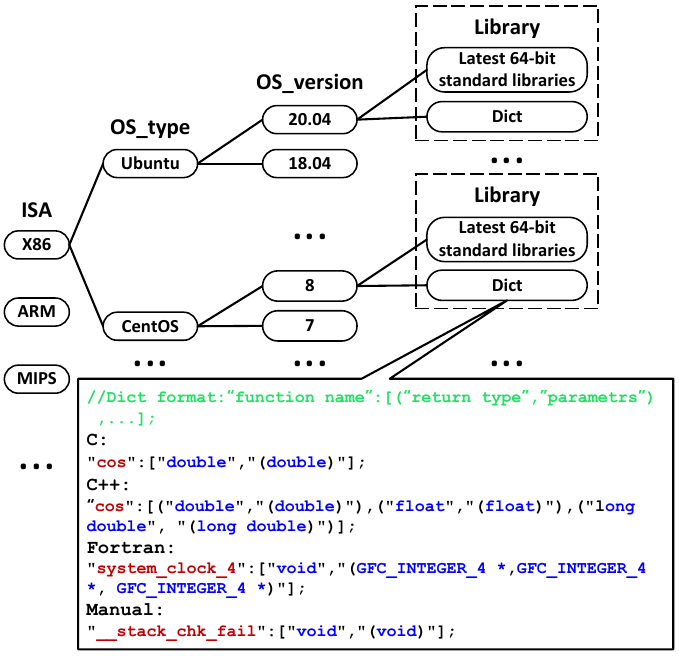}}
\caption{Library Database framework.}
\label{LibraryDatebase}
\end{figure}

In the field of reverse engineering, factors such as ISA, OS, processor architecture, and library version are critical and must be carefully considered for the input ELF. We build a Library Database to cover them. Fig.~\ref{LibraryDatebase} shows the framework of our Library Database, organized layer by layer from ISA, to \textit{OS\_Type}, and finally \textit{OS\_Version}. We do not categorize processor architecture and library versions as separate layers because most ISAs and standard libraries generally offer backward compatibility. (ARM performs poorly in this aspect. We have tried to build a 32-bit ARM library, but the 32-bit ARM is too segmented).
 
\begin{figure}
\centerline{\includegraphics[width=\linewidth]{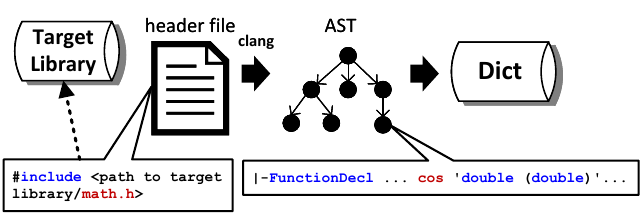}}
\caption{Workflow of the \textit{Dict} auto-generator.}
\label{GenerateDict}
\end{figure}
There is a \textit{Dict} in each library, which is an auto-generated file containing all the function declarations that can be detected by Clang AST-Dump in an external library. The bottom frame in Fig.~\ref{LibraryDatebase} shows a simple example of a \textit{Dict}. It contains the function declarations of EXFs, along with a manually added part. The \textit{Dict} serves three important roles: First, it is the key component for EFACT in generating the output in C/C++ program format; Second, it plays a significant role in our MFC algorithm (introduced in Section \ref{MFC Solver}); Third, despite our efforts to maximize the use of the standard compiler toolchain, certain exceptional cases remain unaddressed. The \textit{Dict} can help us better address these problems. For example, in the bottom frame of Fig.~\ref{LibraryDatebase}, there is a manually added function ``\textit{\_\_stack\_chk\_fail}''. ``\textit{\_\_stack\_chk\_fail}'' is an EXF not adequately structured in the standard library–its declaration and implementation are both within a ``.c'' file, lacking clear separation (the declaration is not in a header file). Since EFACT's auto-completion methodology is traversing the function declarations in the header file suits (as introduced in Section \ref{EFCP}), as a result, EFACT fails to detect this function. To solve this, EFACT incorporates a manual insertion approach. Manual insertion is a common method used in reverse engineering, and Box64~\cite{box64} also employs the same approach. Fortunately, most items appear on the \textit{Dict} can be automatically generated with the method introduced in the next paragraph, and only rare cases trigger the manual insertion mechanism (only 1\%  in SPEC CPU® 2017's EXFs, building on x86-64, Ubuntu 20.04 with GCC 9.4). We take care of those rare manual insertions when building EFACT, thus it's transparent to typical EFACT users.
\begin{figure*}
\centerline{\includegraphics[height=8cm]{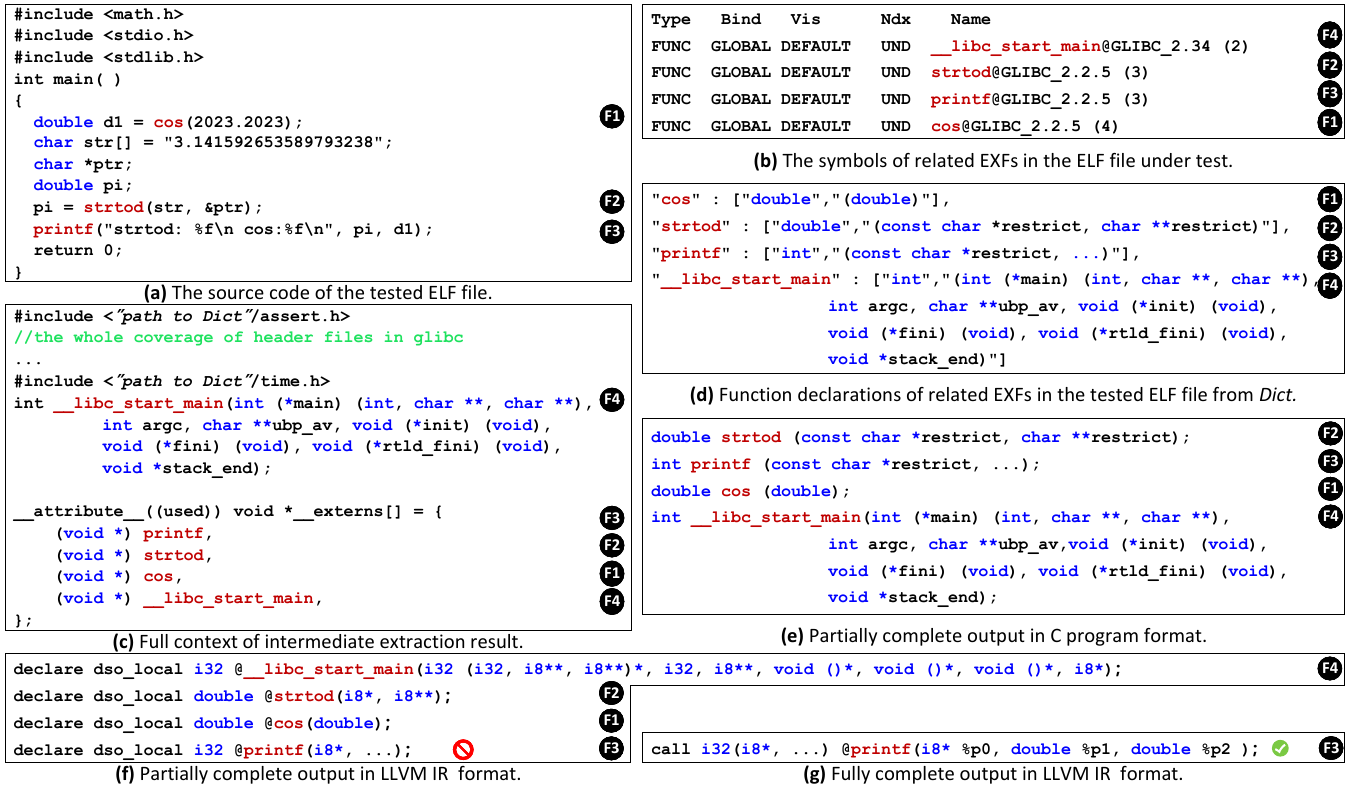}}
\caption{An auto-completion workflow for an ELF compiled from a C program (\textit{This figure is a detailed implementation of the workflow depicted in Fig.~\ref{frameworkovw}, detailed explanation is described in Section \ref{CAutoCompleteSetUp}. The source code of the binary is unknown in actual scenarios, we put it here to better explain our workflow}).}
\label{cAutoComple}
\end{figure*} 
Fig.~\ref{GenerateDict} shows the workflow of the \textit{Dict} auto-generator. First, the \textit{Dict} auto-generator includes header files that define the target library function interfaces into a single file (``header file'' in Fig.~\ref{GenerateDict}), then it uses Clang AST-Dump to generate the abstract syntax tree (AST) of the ``header file''. Afterward, the \textit{Dict} auto-generator traverses the AST, locating the \textit{FunctionDecl}, which refers to a function declaration or definition. Finally, the \textit{Dict} auto-generator records all the \textit{FunctionDecl} and gathers them in a Python \textit{dict} class.
\section{External function completion phase}\label{EFCP}
After choosing a suitable library and obtaining the EXF list, EFACT moves to the completion phase. As mentioned before, there are three main challenges in the completion phase: MFC, FPC, and VPC. When dealing with FPC and VPC, binaries compiled from C/C++ source programs share the same solution approaches. MFC is a special case for C++. In this section, we will introduce the completion workflow for C and C++ separately. Our tool has two formats of completion output: C/C++ program code or LLVM IR. The output in LLVM IR format is the best-recommended format, as it covers all three challenges and maximizes the use of the Clang toolchain. Output in the C/C++ program format does not cover the VPC problem due to the lack of program context.

\subsection{Auto-completion of binaries compiled from C source program}\label{CAutoCompleteSetUp}
Fig.~\ref{cAutoComple} shows an example of the auto-completion workflow for an ELF file compiled from a C source program. This figure is a detailed explanation of the workflow depicted in Fig.~\ref{frameworkovw}. Fig.~\ref{cAutoComple} (a) displays the source code of the tested ELF file. This source code is unknown in practical scenarios, we present it here to facilitate a clearer understanding of the workflow. Fig.~\ref{cAutoComple} (b) illustrates the symbol table of the test ELF file. This symbol table is the initial information that EFACT encounters in step (1) of Fig.~\ref{frameworkovw}. Fig.~\ref{cAutoComple} (c) depicts an intermediate file created by EFACT, which comprises the EXFs extracted from the tested ELF file and the \textit{Dict} selected by EFACT to aid in the auto-completion of these EXFs. This intermediate file results from the completion of steps (1), (2), and (3) in Fig.~\ref{frameworkovw}, as input to the Func Auto-Completer. Fig.~\ref{cAutoComple} (d) shows the function declarations of the relevant EXFs from our chosen \textit{Dict}. Fig.~\ref{cAutoComple} (e) presents the partially complete output in C program format, and Fig.~\ref{cAutoComple} (f) displays the partially complete output in LLVM IR format, both corresponding to the outcomes of step (6) in Fig.~\ref{frameworkovw}. Lastly, Fig.~\ref{cAutoComple} (g) exhibits the fully complete output in LLVM IR format, aligning with the results of step (7) in Fig.~\ref{frameworkovw}. There are four EXFs in this example, which we treat as F1, F2, F3, and F4. F1, F2, and F4 belong to the FPC problem, while F3 belongs to the VPC problem. From the source code, it is apparent that F4 is not explicitly used. This type of function, always called by the OS and inaccessible to programmers directly, is one of the functions we declared ``manually'' in Section \ref{librarydatabases}.

\subsubsection{\textbf{FPC Solver}}\label{FPC SOLVER}
For output in LLVM IR format, our method involves putting the function name into a list, as shown in Fig.~\ref{cAutoComple} (c), including the target library's header file, and then passing this file to the Clang with the command (``\textit{complete.c}'' corresponds to Fig.~\ref{cAutoComple} (c) and ``\textit{llvm\_ir\_output.bc}'' corresponds to Fig.~\ref{cAutoComple} (f)):
 
\textit{clang -emit-llvm -S complete.c -o llvm\_ir\_output.bc.}\\In the ``\textit{complete.c}'' file, EFACT automatically adds the full declaration of ``\textit{\_\_libc\_start\_main}'' (as defined \textit{manually} in Section \ref{librarydatabases}) to avoid compile errors. Fig.~\ref{cAutoComple} (f) shows the recovered function declarations in LLVM IR format output from Clang, where we can see that F1, F2, and F4 have the correct declarations. However, this method cannot complete \textit{printf}. EFACT leaves it to our VPC Solver. EFACT automatically generates the output in C program format with the information in \textit{Dict}.
\subsubsection{\textbf{VPC Solver}}\label{VPC SOLVER}
Different from fixed-parameter functions, variable-parameter functions, like \textit{printf}, do not have a certain declaration. We see ``...'' (introduced in Section \ref{THREE CHALLANGE}) in the function declaration of \textit{printf}. These parameters depend on the ``\textit{\_\_restrict \_\_fmt}'' format passed to \textit{printf}. From a reverse-engineering standpoint, merely having the definition available in the standard library is insufficient for function declaration recovery; the source binary's context is vital. To address this, we developed an LLVM Pass, \textit{PrintfPass} to facilitate this functionality. This pass operates on the lifted IR context of the application binary, generated via McSema. We will introduce it based on the example in Fig.~\ref{cAutoComple} (a).
 
Firstly, \textit{PrintfPass} locates the associated string context of ``\textit{\_\_restrict \_\_fmt}''. As shown in Fig.~\ref{cAutoComple} (a), the associated string is ``\textit{strtod: \%f $\setminus$n cos:\%f $\setminus$n}''. Secondly, \textit{PrintfPass} conducts a syntactic analysis of this string. \textit{PrintfPass} employs regular expressions to sequentially match format specifiers like ``\textit{\%f}''. These format specifiers within the string dictate the necessary corresponding parameters to be included as arguments in the \textit{printf} function. Thirdly, upon matching these format specifiers, our \textit{PrintfPass}, following the detected platform details and the standard guidelines defined in the \textit{printf} function documentation\footnote{https://cplusplus.com/reference/cstdio/printf/}, replaces the specifiers and recovers the ``...'' part of \textit{printf} with the actual parameters. Fig.~\ref{cAutoComple} (g) displays the outcome of this process. The same approach can be easily adapted to other variable-parameter functions, such as \textit{scanf}.

\begin{figure}
\centerline{\includegraphics[width=\linewidth]{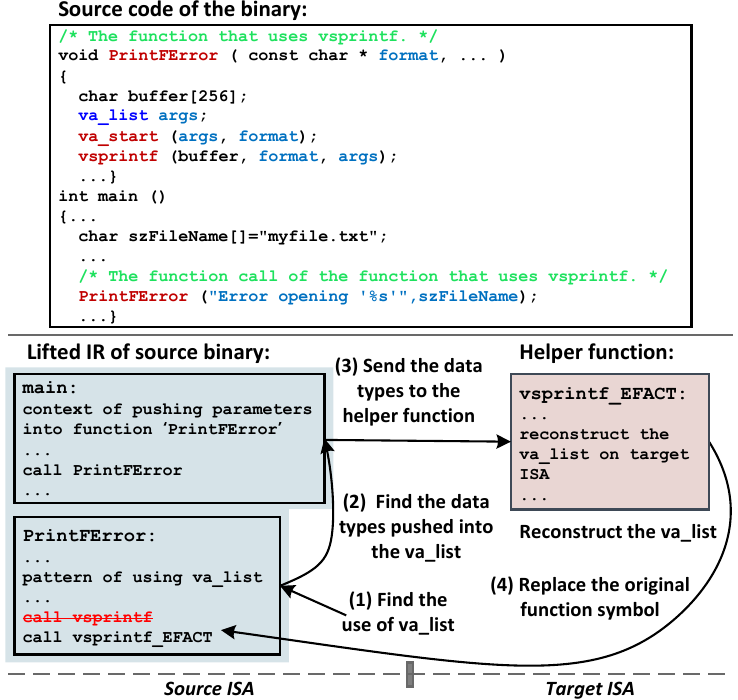}}
\caption{Workflow of cross-ISA \textit{va\_list} translation based on \textit{Va\_listPass} (source code of the binary is unknown in actual scenarios, we put it here to better explain our workflow).}
\label{cross-ISA}
\end{figure}
 
As mentioned in Section \ref{THREE CHALLANGE}, when introducing VPC, the variable parameter represented with \textit{va\_list} needs special handling during cross-ISA static binary translation. \textit{va\_list} is a type widely used in C/C++ to manage variable argument lists, prominently utilized by functions like \textit{vsprintf}, \textit{vfprintf}, etc. Its initiation method is the same on x86 and ARM. However, its specific implementation on the stack differs between the two. Reconstructing the physical stack frames of the source ISA on the target ISA stands as a pivotal research problem in static binary translation. Misalignment of data on the stack can result in corrupted translated binaries. To address this, we develop another LLVM Pass named \textit{Va\_listPass}, the workflow of which is depicted in Fig.~\ref{cross-ISA}. To illustrate the functioning of \textit{Va\_listPass}, we use \textit{vsprintf} as an example. The upper half of Fig.~\ref{cross-ISA} demonstrates a typical usage scenario of \textit{vsprintf} and \textit{va\_list}. The \textit{va\_list} is initialized using \textit{va\_start}. The ``\textit{format}'' parameter in \textit{va\_start} and the ``\textit{args}'' parameter in \textit{vsprintf} are associated with the parameters passed from the function that utilizes them. The ``\textit{format}'' in \textit{vsprintf} serves a similar purpose (format specifiers, which dictate the necessary corresponding parameters to be included as arguments) as ``\textit{\_\_restrict \_\_fmt}'' does in \textit{printf}. To recover the \textit{vsprintf}, \textit{Va\_listPass} undertakes the following operations:
 
\textbf{(1)} Scan the lifted IR of the source binary and locate the invocation of \textit{va\_start}. \textit{va\_start} is the detailed implementation start of \textit{va\_list}.
 
\textbf{(2)} Detect the invocation of the function that utilizes \textit{va\_list}, such as \textit{vsprintf} here, and employs a method akin to \textit{PrintfPass} (as introduced in the second paragraph of Section \ref{VPC SOLVER}). This method is used to determine the data types of the parameters that are subsequently pushed into the \textit{va\_list}.
 
\textbf{(3)} Constructs a helper function on the target ISA, accepting the data types identified in step (2) as parameters. This helper function facilitates the correct reconstruction of \textit{va\_list} on the target ISA during execution.
 
\textbf{(4)} Replaces the original function call of \textit{vsprintf} with the new helper function. 
 
Following this process, we ensure that \textit{va\_list} can be accurately translated across multiple ISAs. Besides, \textit{std::format} is a new function introduced in C++ 20, it is more efficient than previous string formatting options in C++, such as \textit{vsprintf} and \textit{printf}. With the help of the MFC algorithm mentioned in the next paragraph, the way we treat \textit{vsprintf} and \textit{printf} can be transplanted to \textit{std::format}.
 
To be noticed, not all variable-parameter functions follow a format parameter similar to \textit{printf} for determining their specific parameter lists. Attaining comprehensive coverage for all variable-parameter functions presents a considerable challenge. Based on our engineering experience, the approach to recovering variable-parameter functions often varies from one case to another. In developing the VPC Solver of EFACT, we prioritized practical applications, initially focusing on supporting commonly encountered variable-parameter functions. To elaborate, the current version of EFACT’s VPC Solver is tailored based on the variable-parameter functions observed in EEMBC and SPEC CPU\circledR 2017 benchmarks. We believe that covering these benchmarks showcases the practical utility of our tool.

\subsection{Auto-completion of binaries compiled from C++  source program}\label{C++autocomple}
Fig.~\ref{C++AutoComple} provides a detailed breakdown of the workflow for recovering an ELF file compiled from a C++ source program, as outlined in Fig.~\ref{frameworkovw}. Apart from the challenges associated with name mangling, EFACT's methodology for handling C++ is consistent with the approaches used for C. Consequently, Fig.~\ref{C++AutoComple} focuses specifically on the recovery of mangled functions. Fig.~\ref{C++AutoComple} (a) displays the output from LLVM's Demangling API. Fig.~\ref{C++AutoComple} (b) illustrates the matched return types of the mangled function from our chosen \textit{Dict}. Fig.~\ref{C++AutoComple} (c) details the specific additions made to the intermediate file (as mentioned in Section \ref{CAutoCompleteSetUp}) for handling mangled functions. Fig.~\ref{C++AutoComple} (a), (b), (c) are all generated in step (4) of Fig.~\ref{frameworkovw}. Fig.~\ref{C++AutoComple} (d) presents the output in C++ program format and LLVM IR format, corresponding to the outcomes of step (6) in Fig.~\ref{frameworkovw}. Additionally, Fig.~\ref{C++AutoComple} (d) shows the output from RetDec for comparison. In this section, our discussion begins with an introduction to LLVM’s Demangling API, followed by an explanation of how we address the name mangling problem.
\subsubsection{\textbf{LLVM’s Demangling API and MFC Solver}}\label{MFC Solver}
\begin{figure}
\centerline{\includegraphics[height=11.5cm]{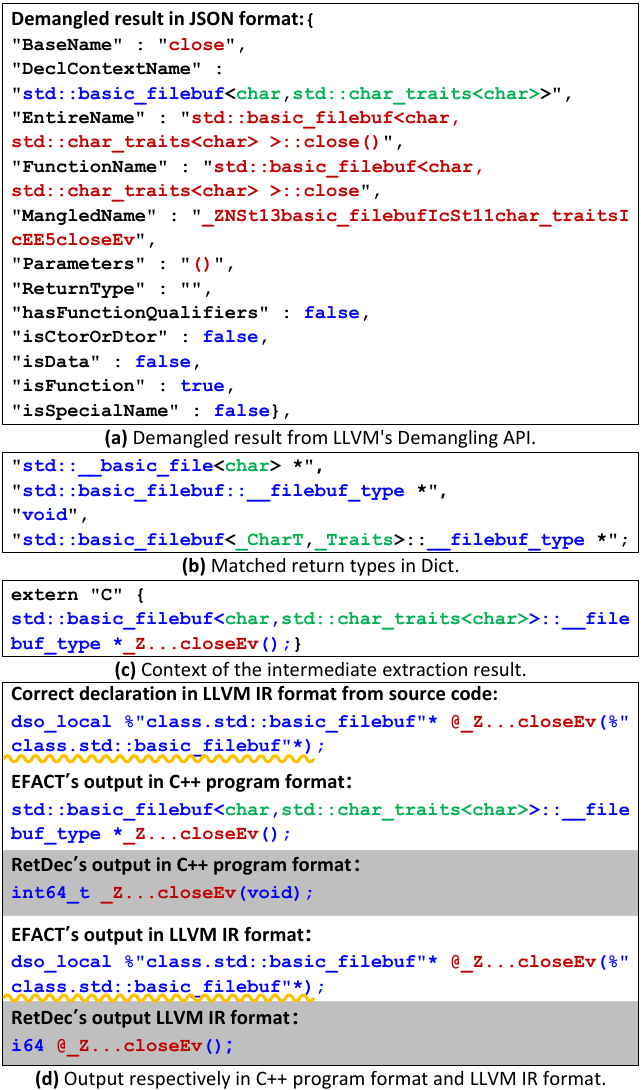}}
\caption{An auto-completion workflow for an ELF compiled from a C++ program (\textit{This figure is a detailed implementation of the workflow depicted in Fig.~\ref{frameworkovw}, a detailed explanation is described in Section \ref{C++autocomple}}).} 
\label{C++AutoComple}
\end{figure} 
We can see ``\textit{\_ZN...@GLIBCXX\_3.4}'' in the symbol table of a C++ program. From the keyword ``\textit{@GLIBCXX}'' we know this is a standard function defined in \textit{libstdc++}. However, a search through the \textit{libstdc++’s} source code fails to yield this function unless we demangle the EXF, showcasing a typical instance of C++ name mangling. Fig.~\ref{C++AutoComple} shows an example of how EFACT handles the mangled function. First, EFACT directly uses the LLVM’s Demangling API to demangle the symbol. From Fig.~\ref{C++AutoComple} (a), we see the output from the LLVM's Demangling API, the original function name is ``\textit{close}'', which is a widely used function in C++. Second, EFACT reconstructs the demangled result to a C++ function, shown in Fig.~\ref{C++AutoComple} (c). We use the \textit{extern ``C''} linkage specification to let the compiler not apply name mangling to this function. Then, EFACT passes it to the compiler and gets the completed result, which is shown in Fig.~\ref{C++AutoComple} (d).
 
In the portion highlighted by the yellow wavy line in Fig.~\ref{C++AutoComple} (d), we observe an additional parameter compared to the ``Parameters'' obtained from the LLVM's Demangling API. This extra parameter represents the implicit \textit{this} pointer, a fundamental element in C++. Existing tools (like RetDec in Fig.~\ref{C++AutoComple} (d)) all omit the implicit \textit{this} pointer in their lifting outputs of LLVM IR. They seem to overlook that the output from the LLVM's Demangling API is primarily designed for C++, and thus a more profound adaptation is necessary when lifting to LLVM IR. Omission of the \textit{this} pointer not only misrepresents the source binary but also risks inducing parameter discrepancies during static binary translation, potentially leading to execution errors or program crashes.
 
From Fig.~\ref{C++AutoComple} (a), we can see the ``Return Type'' of this function is empty. Although the LLVM’s Demangling API reliably extracts the parameters, it does not guarantee the retrieval of the return type. Existing tools fall short of thoroughly investigating this aspect. As shown in Fig.~\ref{C++AutoComple} (d), they simply treat all the unknown cases with an \textit{int} (in C/C++) or \textit{i32}/\textit{i64} (in LLVM IR). This approach proves to be inadequate, especially in scenarios where the actual return type is \textit{double}, as exemplified in Fig.~\ref{motivatingExample} (c). Following this procedure can result in incorrect lifting outcomes, ultimately corrupting the program after static binary translation. Besides, EFACT's output recovers more context information of the EXF than RetDec.
 
\begin{figure}
\centerline{\includegraphics[height=6.2cm]{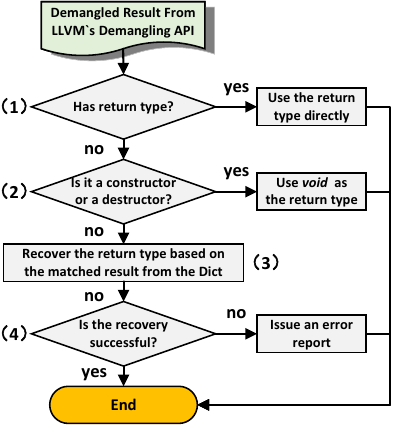}}
\caption{Workflow of the MFC algorithm.}
\label{MFC}
\end{figure}
The main reason stopping existing tools from going deeper is the template specialization mechanism in C++. As shown in Fig.~\ref{C++AutoComple} (b), when we try to extract the standard return type from the header file, we might encounter returns that are embedded with templates, necessitating further context-specific adjustments. Our MFC algorithm draws from general specifications for function declarations in C++ and implements the return value from LLVM’s Demangling API. EFACT also utilizes the previously mentioned \textit{Dict} to supplement the missing information from the LLVM’s Demangling API, enhancing the accuracy and comprehensiveness of our approach. Fig.~\ref{MFC} illustrates the workflow of our MFC algorithm, our algorithm has the following steps:
 
\textbf{(1)} Checking the demangled result from the LLVM's Demangling API, if the ``ReturnType'' is not empty, we use it directly.

\textbf{(2)} When the ``ReturnType'' is empty, we inspect the ``isCtorOrDtor'' flag. This flag helps in identifying whether the function is a constructor or a destructor. Given that the constructor and destructor always return a \textit{void}, we use \textit{void} as a return type.

\textbf{(3)} If ``ReturnType'' and ``isCtorOrDtor'' all failed, we utilize the "BaseName" and "Parameters" retrieved from the LLVM’s Demangling API, cross-verifying this data against the matched result from the \textit{Dict}. If there's a single matched return type that doesn't contain templates, we select it as the result. If multiple matched returns or templates are identified, we pinpoint the correct return type based on the ``DeclContextName''. This item contains detailed template information, assisting us in identifying the correct results. For example, in Fig.~\ref{C++AutoComple} (b), function ``\_ZNSt...closeEv'' has four matched returns in \textit{Dict}. From the ``DeclContextName'', we can see this function is declared from \textit{std::basic\_filebuf}, with two facilities templates. This context allows us to correctly match it with the fourth result, and replace all the templates in the result with the actual entities specified in ``DeclContextName'', as highlighted in the green part in Fig.~\ref{C++AutoComple} (c). 

\textbf{(4)} If the aforementioned steps prove unsuccessful, EFACT will issue an error report, and we recommend manually adding the function.

Fig.~\ref{C++AutoComple} (d) shows a comparison of the recovered results from EFACT and RetDec. We can see our recovery preserves more contextual information about the original function, which aids in a more comprehensive understanding of the source binary.
\subsection{Special case}
\begin{figure}
\centerline{\includegraphics[height=3.3cm]{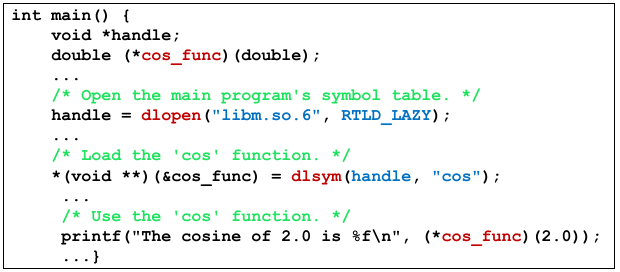}}
\caption{An example of \textit{dlopen} and \textit{dlsym}.}
\label{dlopen}
\end{figure}
Sometimes, the function symbol of an EXF might not be present in the symbol table of the source ELF file. For instance, as shown in Fig.~\ref{dlopen}, when ``\textit{cos}'' is loaded using \textit{dlopen} and \textit{dlsym}, ``\textit{cos}'' doesn't appear in the symbol table. The recovery process in such a case necessitates the context of the source binary. To tackle this challenge, we have developed another LLVM Pass, named \textit{DlopenPass}. The process of \textit{DlopenPass} involves the following steps:
 
\textbf{(1)} Scan the entire lifted IR of the source binary and locate all the invocations of \textit{dlopen} and \textit{dlsym}.
 
\textbf{(2)} Find the first parameter passed to \textit{dlopen} to get the name of the target shared library. If this library is not included in EFACT’s support libraries, generate an error report.

\textbf{(3)} If the target shared library is included in EFACT’s support libraries, the next step is to identify the second parameter passed into \textit{dlsym} and ascertain the corresponding function name.
 
\textbf{(4)} Add this function to the external function list and recover this function through EFACT's standard workflow.

It's important to note that the parameters examined in steps (2) and (3) may not always be constant strings. In such a case, determining the exact names of the target shared libraries or loaded functions becomes challenging. This requires additional dataflow analysis to figure out these names, thus expanding the coverage for more complicated use cases of \textit{dlopen}/\textit{dlsym}. We plan to incorporate this advanced feature in the future version of EFACT. Fortunately, our current solution can resolve many typical use cases of \textit{dlopen}/\textit{dlsym}.

\section{Evaluation}\label{Evalua}
\subsection{Evaluation setup}
To evaluate the capability of EFACT, we select five key dimensions for evaluation. In this section, we will explain the reason behind choosing each of these dimensions and describe the specific experiment designed to evaluate their respective impacts:
 
\textbf{Coverage: }This dimension is chosen to evaluate EFACT's capability to handle ELF files from diverse platforms. In reverse engineering, it is common to encounter ELF files built on numerous different platforms. This evaluation is presented as \textbf{Experiment 1} in Section \ref{CoverageExp}. The methodology for experiment 1 involves using standard benchmarks to test whether EFACT can recover EXFs in these benchmarks. To simulate complex real-world workloads, we employ benchmarks from SPEC CPU® 2017. And this benchmark is constructed considering various factors, as detailed in Table \ref{tab2}, such as \textit{OS\_type} (\textit{HD\_x86} vs \textit{VM\_x86\_1}), \textit{OS\_version} (\textit{HD\_x86} vs \textit{VM\_x86\_2}) and ISA (\textit{HD\_x86} vs \textit{HD\_ARM}). The four hardware configurations listed in Table \ref{tab2} are utilized to build the test benchmarks, aiming to replicate the complexity of real-world workloads as closely as possible. EFACT itself runs on \textit{HD\_x86}, which also proves its capability of lifting binaries built for different platforms.
\begin{table}
\centering
\caption{Experimental setup (GCC 9.4).}
\renewcommand{\arraystretch}{1.2}
\resizebox{0.48\textwidth}{!}{
\begin{tabular}{llll}

\hline
\textbf{Name}&\textbf{OS} &\textbf{ISA}&\textbf{CPU}\\
\cline{1-4}
HD\_x86&Ubuntu 20.04&x86-64&{Intel® Xeon® Gold 5218R}\\
HD\_ARM&Ubuntu 20.04&AArch64&{Ampere® Altra®}\\
VM\_x86\_1&Centos 8&x86-64&{QEMU Virtual CPU}\\
VM\_x86\_2&Ubuntu 18.04&x86-64&{QEMU Virtual CPU}\\
\hline
\end{tabular}
\label{tab2}
}
\end{table}
\textbf{Correctness: }Merely producing a recovered output does not guarantee its accuracy. It is essential to evaluate the correctness of the tool, ensuring that the recovered results match the original function declarations in the standard library. We evaluate this dimension as \textbf{Experiment 2} in Section \ref{CorrectnessExp}. The methodology for experiment 2 involves a manual comparison to ascertain whether EFACT's recovered results align with the corresponding declarations in the standard library. We compare EFACT's performance in this dimension with state-of-the-art static binary lifting tools like RetDec and McSema. Recognizing that manual checking is time-consuming, we opted for a smaller set of evaluation inputs compared to experiment 1. For this experiment, we selected SPEC CPU® 2017, built on \textit{HD\_x86} with \textit{64-bit} processor architecture, and used the \textit{O3} optimization option, to assess the correctness of EFACT's output. Experiment 2 focuses particularly on the correctness of recovery in mangled functions, as this is a key area of contribution from us.

\textbf{Extensibility: }As discussed in Section \ref{EFCPfRAME}, a framework only for C/C++ may fall short in addressing the diverse requirements encountered in actual scenarios. EFACT is designed to support lifting binaries compiled from languages other than C/C++ and integrating libraries beyond \textit{glibc}/\textit{libstdc++}. We perform \textbf{Experiment 3} in Section \ref{ExtensibilityExp} to evaluate this dimension. The evaluation approach of experiment 3 uses benchmarks whose source languages are not C/C++ as inputs for EFACT. Additionally, we employ an ELF file that utilizes functions from libraries other than \textit{glibc}/\textit{libstdc++} to ascertain EFACT's support for diverse libraries. We select the Fortran benchmarks from SPEC CPU® 2017, along with 10 command-line utilities written in Rust~\cite{RustCommand}, to evaluate EFACT's extensibility. As a test case for a new library, we choose \textit{libgfortran} and OpenSSL.
 
\textbf{Usability: }As discussed in Section \ref{introduciton}, current static binary lifting tools have limitations in solving the EXFC problem. Our goal is to enhance their capability in addressing EXFC more easily. We perform \textbf{Experiment 4} in Section \ref{UsabilityEXP} to evaluate the usability of EFACT. The methodology for experiment 4 involves assessing whether our tool can aid existing tools in more effectively lifting the input binary. We have chosen Lasagne and McSema as the target tools for enhancement and selected EEMBC as the evaluation input. In Section \ref{UsabilityEXP}, we do not include the evaluation for McSema directly. Instead, we focus on EFACT\_MC, a static binary translator that integrates EFACT with McSema. The evaluation of EFACT\_MC can demonstrate how EFACT can assist McSema in improving binary lifting. The details of this evaluation will be elaborated in Section \ref{SBTEXP}.

\textbf{Static binary translation: }We have invested a lot of effort in the EXFC required by static binary translation. To evaluate our effort, we developed a static binary translator named EFACT\_MC. This tool integrates EFACT with McSema to facilitate binary translation from x86-64 to AArch64. EFACT\_MC initially lifts an x86-64 binary to LLVM IR and subsequently employs the LLVM compiler to generate the target binary. We evaluate the performance of EFACT\_MC in Section \ref{SBTEXP} as \textbf{Experiment 5}. The methodology of experiment 5 involves assessing whether EFACT\_MC achieves higher translation accuracy compared to McSema alone. For this purpose, we utilize EEMBC as the evaluation input.  Initially, we attempted to use SPEC CPU® 2017 as test input but encountered numerous errors (such as register mapping) while employing McSema to lift. These errors are not directly related to our primary focus, and addressing them would demand significant effort. Consequently, we use EEMBC, a benchmark that has a smaller code size but is nearly as authoritative to evaluate EFACT\_MC. While the current version of EFACT\_MC primarily serves to evaluate the EXFC accuracy of our framework, there is an ongoing commitment to further refine and enhance its overall performance in future developments.

\textbf{Beyond the aforementioned five dimensions,} we also characterize the SPEC CPU® 2017 from the perspective of EXF. This analysis is presented as \textbf{Experiment 6} in Section \ref{EXFEXP}. The purpose of this experiment is to assist binary lifting researchers in gaining a deeper understanding of the significance of EXFC. This experiment also reveals the factors that impact the generation of EXFs.
 
RetDec (v5.0), Lasagne (v3\footnote{https://doi.org/10.5281/zenodo.6408463}), and McSema (v3.0.26) are used in our experiments, and their versions are the latest ones when we conducted our research. For all benchmarks, except those in Table \ref{tab10} and Table \ref{tab12}, we employ the \textit{O3} optimization option. This option is the most commonly used setting in the best-known configuration (BKC) for these benchmarks. For the benchmarks in Table \ref{tab10} and Table \ref{tab12}, we use the \textit{O0} optimization option to minimize the compiler's impact on different ISAs and processor architectures. 
 
Current Library Database is built with the latest version (when we conducted our research) of \textit{glibc} (v2.37), \textit{libstdc++} (v3.4.30), \textit{libgfortran} (v10) and OpenSSL (v3.1.1). EFACT identifies EXFs in \textit{glibc} using the symbol “\textit{@GLIBC}”, in \textit{libstdc++} with “\textit{@GLIBCXX}”, in \textit{libgfortran} with “\textit{@GFORTRAN}” and in OpenSSL with “\textit{@OPENSSL}”. We aim to expand the range of symbols recognized in future versions of EFACT. SPEC CPU\circledR 2017 is organized into 4 suites: SPECrate 2017 Integer (intrate), SPECspeed 2017 Integer (intspeed), SPECrate 2017 Floating Point(fprate), and SPECspeed 2017 Floating Point (fpspeed).

\subsection{Coverage and correctness}
\subsubsection{Coverage}\label{CoverageExp}
\begin{table}
\centering
\caption{EFACT's coverage on EXFs from \textit{glibc}, \textit{libstdc++} and \textit{libgfortran} in SPEC CPU\circledR 2017 (\textit{O3}).}

\renewcommand{\arraystretch}{1.2}
\resizebox{0.35\textwidth}{!}{
\begin{tabular}{lcc}

\hline
\textbf{Platform}&\textbf{EXF amount}&\textbf{Coverage}  \\
\hline
HD\_x86&3679&100\%  \\
HD\_ARM&3715&100\%  \\
VM\_x86\_1&3644&100\%  \\
VM\_x86\_2&3694&100\%  \\

\hline
\multicolumn{3}{l}{\textit{*Definition of Coverage is shown in definition \ref{definition1}.}}\\
\end{tabular}
\label{tab3}

}
\end{table}
We use SPEC CPU\circledR 2017 to evaluate the coverage of our tool. The ability of coverage is defined as follows:
\newdefinition{Definition}{Definition}
\begin{Definition}
Coverage = (Number of EXFs can be completed by EFACT) / (Number of EXFs in the target ELF file from \textit{glibc}, \textit{libstdc++} and \textit{libgfortran})
\label{definition1}
\end{Definition}
Table \ref{tab3} shows the coverage of EFACT. From the ``EXF amount'' column, we can see the platform does impact the generation of EXFs. First, when examining different ISAs, we observe that the EXF amount on \textit{HD\_ARM} exceeds that on \textit{HD\_x86}. This difference is attributed to ARM's weak memory model and RISC architecture, which we have detailed discussed in Section \ref{EXFEXP}. Second, when facing different OS, the EXF amounts on \textit{Ubuntu 20.04} (\textit{HD\_x86}) and \textit{Ubuntu 18.04} (\textit{VM\_x86\_2}) are higher than \textit{CentOS 8} \textit{(VM\_x86\_1}). On Ubuntu systems, compiling with the \textit{O3} optimization option triggers the ``\_FORTIFY\_SOURCE'' macro to be set to ``2'', leading to the generation of many EXFs for buffer overflow detection. Conversely, on \textit{CentOS 8},  the ``\_FORTIFY\_SOURCE'' macro is not defined, and the compiler does not enforce buffer overflow detection. Concerning different versions of the same OS, we note that the EXF amount on \textit{VM\_x86\_2} is bigger than the amount on \textit{HD\_x86}, with the additional EXFs being functions related to buffer overflow detection. EFACT can achieve 100\% coverage on the EXFs from \textit{glibc}, \textit{libstdc++}, and \textit{libgfortran} of all benchmarks, not restricted by the platform.
\subsubsection{Correctness}\label{CorrectnessExp}
\begin{table}
\centering
\caption{EFACT's detailed analysis result on EXFs from \textit{glibc}, \textit{libstdc++} and \textit{libgfortran} in SPEC CPU® 2017 (\textit{HD\_x86}, \textit{64-bit}, \textit{O3}).}

\renewcommand{\arraystretch}{1.2}
\resizebox{0.41\textwidth}{!}{
\begin{tabular}{lcccccc}

\hline
 \multirow{2}*{\textbf{Suite}}&\multirow{2}*{\textbf{EXFs}}&\multirow{2}*{\textbf{VPC}}&\multirow{2}*{\textbf{FPC}}&\multicolumn{3}{c}{\textbf{MFC}} \\
\cline{5-7}
&&&&\multicolumn{1}{c}{\textbf{num}}&\multicolumn{1}{c}{\textbf{this}}&\multicolumn{1}{c}{\textbf{has ret}} \\

\hline
intspeed&\multicolumn{1}{c}{804}&\multicolumn{1}{c}{51}&\multicolumn{1}{c}{593}&\multicolumn{1}{c}{160}&\multicolumn{1}{c}{119}&\multicolumn{1}{c}{21}\\
intrate&\multicolumn{1}{c}{800}&\multicolumn{1}{c}{51}&\multicolumn{1}{c}{591}&\multicolumn{1}{c}{158}&\multicolumn{1}{c}{117}&\multicolumn{1}{c}{18}\\
fpspeed&\multicolumn{1}{c}{847}&\multicolumn{1}{c}{43}&\multicolumn{1}{c}{760}&\multicolumn{1}{c}{44}&\multicolumn{1}{c}{29}&\multicolumn{1}{c}{5}\\
fprate&\multicolumn{1}{c}{1228}&\multicolumn{1}{c}{62}&\multicolumn{1}{c}{993}&\multicolumn{1}{c}
{173}&\multicolumn{1}{c}{107}&\multicolumn{1}{c}{28}\\

\cline{1-7}
\textbf{Ratio}&&&&&\multicolumn{1}{c}{69.5\%}&\multicolumn{1}{c}{13.5\%}\\

\hline
\multicolumn{7}{l}{\textit{*\textbf{this}: num of functions which has a ``this'' pointer as para-}}\\
\multicolumn{7}{l}{\textit{meter. \textbf{has ret:} num of demangled outputs from LLVM's}}\\
\multicolumn{7}{l}{\textit{Demangling API that has a return type. SPEC CPU® 20-}}\\
\multicolumn{7}{l}{\textit{17 is organized into 4 suites.} \textbf{MFC num} $\neq$ \textbf{this} + \textbf{has ret}.}\\
\end{tabular}
\label{tab4}

}
\end{table}
\begin{table}
\centering
\caption{Comparation among RetDec, McSema, and EFACT on the correctness of EXFC from \textit{glibc}, \textit{libstdc++} and \textit{libgfortran} on SPEC CPU\circledR 2017 (LLVM IR format, \textit{HD\_x86}, \textit{O3}, \textit{64-bit}).}

\renewcommand{\arraystretch}{1.2}
\resizebox{0.45\textwidth}{!}{
\begin{tabular}{lcccc}

\hline

\multirow{2}*{\textbf{Function Type}}&{\textbf{EXF}}&\multicolumn{3}{c}{\textbf{Correctness}} \\
\cline{3-5}
&\textbf{amount}&\multicolumn{1}{c}{\textbf{McSema}}&\multicolumn{1}{c}{\textbf{RetDec}}&\multicolumn{1}{c}{\textbf{EFACT}} \\

\hline
variable parameter&120&0\%&100\%&100\%  \\
fixed parameter&1612&4.2\%&89.7\%&100\%  \\
mangled&444&1.3\%&4.7\%&100\%  \\

\hline
\multicolumn{5}{l}{\textit{*Definition of Correctness is shown in definition \ref{definition2}}.}\\
\end{tabular}
\label{tab5}
}
\end{table}
\begin{table*}
\caption{Comparation among RetDec, McSema, and EFACT on the ability of MFC on SPECrate 2017 (LLVM IR format, \textit{HD\_x86}, \textit{O3}, \textit{64-bit}).}

\renewcommand{\arraystretch}{1.2}
\resizebox{1\textwidth}{!}{
\begin{tabular}{lrrrcccccccccrrrr}

\hline
\multirow{2}*{\textbf{Benchmark}}&\multicolumn{1}{c}{\textbf{EXF}}&\multicolumn{1}{c}{\textbf{info}}&\multicolumn{4}{c}{\textbf{RetDec}}&&\multicolumn{4}{c}{\textbf{McSema}}&&\multicolumn{4}{c}{\textbf{EFACT}}\\
\cline{4-7}\cline{9-12}\cline{14-17}
&\multicolumn{1}{c}{\textbf{num}}&\multicolumn{1}{c}{\textbf{included}}&\multicolumn{1}{c}{\textbf{AC}}&\multicolumn{1}{c}{\textbf{crash}}&\multicolumn{1}{c}{\textbf{widening}}&\multicolumn{1}{c}{\textbf{info loss}}&&\multicolumn{1}{c}{\textbf{AC}}&\multicolumn{1}{c}{\textbf{crash}}&\multicolumn{1}{c}{\textbf{widening}}&\multicolumn{1}{c}{\textbf{info loss}}&&\multicolumn{1}{c}{\textbf{AC}}&\multicolumn{1}{c}{\textbf{crash}}&\multicolumn{1}{c}{\textbf{widening}}&\multicolumn{1}{c}{\textbf{info loss}}\\
\cline{1-17}
\multicolumn{1}{l}{leela\_r}&\multicolumn{1}{c}{53}&\multicolumn{1}{c}{25} 
&\multicolumn{1}{c}{2} &\colorbox{red}{1}&\multicolumn{1}{c}{50}&\multicolumn{1}{c}{25}
&
&\multicolumn{1}{c}{1} &\multicolumn{1}{c}{0}&\multicolumn{1}{c}{52}&\multicolumn{1}{c}{25}
&
&\multicolumn{1}{c}{53}&\multicolumn{1}{c}{0}&\multicolumn{1}{c}{0}&\multicolumn{1}{c}{0}
\\
\multicolumn{1}{l}{xalancbmk\_r}&\multicolumn{1}{c}{31}&\multicolumn{1}{c}{13}&\multicolumn{1}{c}{0} &\multicolumn{1}{c}{0}&\multicolumn{1}{c}{31}&\multicolumn{1}{c}{13}
&
&\multicolumn{1}{c}{0} &\multicolumn{1}{c}{0}&\multicolumn{1}{c}{31}&\multicolumn{1}{c}{13}
&
&\multicolumn{1}{c}{31}&\multicolumn{1}{c}{0}&\multicolumn{1}{c}{0}&\multicolumn{1}{c}{0} 
\\
\multicolumn{1}{l}{omnetpp\_r}&\multicolumn{1}{c}{76}&\multicolumn{1}{c}{35}&\multicolumn{1}{c}{6} &\colorbox{red}{6}&\multicolumn{1}{c}{64}&\multicolumn{1}{c}{35}   
&
&\multicolumn{1}{c}{4} &\colorbox{red}{2}&\multicolumn{1}{c}{70}&\multicolumn{1}{c}{35}  
&
&\multicolumn{1}{c}{76}&\multicolumn{1}{c}{0}&\multicolumn{1}{c}{0}&\multicolumn{1}{c}{0}
\\
\multicolumn{1}{l}{cactuBSSN\_r}&\multicolumn{1}{c}{44}&\multicolumn{1}{c}{17}&\multicolumn{1}{c}{1} &\colorbox{red}{1}&\multicolumn{1}{c}{42}&\multicolumn{1}{c}{17}  
&
&\multicolumn{1}{c}{1}&\multicolumn{1}{c}{0}&\multicolumn{1}{c}{43}&\multicolumn{1}{c}{17}
&
&\multicolumn{1}{c}{44}&\multicolumn{1}{c}{0}&\multicolumn{1}{c}{0}&\multicolumn{1}{c}{0}
\\
\multicolumn{1}{l}{namd\_r}&\multicolumn{1}{c}{6}&\multicolumn{1}{c}{0}&\multicolumn{1}{c}{0} &\multicolumn{1}{c}{0}&\multicolumn{1}{c}{6}&\multicolumn{1}{c}{0} 
&
&\multicolumn{1}{c}{0} &\multicolumn{1}{c}{0}&\multicolumn{1}{c}{6}&\multicolumn{1}{c}{0}
&
&\multicolumn{1}{c}{6}&\multicolumn{1}{c}{0}&\multicolumn{1}{c}{0}&\multicolumn{1}{c}{0}
\\
\multicolumn{1}{l}{parest\_r} &\multicolumn{1}{c}{101}&\multicolumn{1}{c}{51}&\multicolumn{1}{c}{3} &\colorbox{red}{4}&\multicolumn{1}{c}{94} &\multicolumn{1}{c}{51}
&
&\multicolumn{1}{c}{3} &\colorbox{red}{1}&\multicolumn{1}{c}{97} &\multicolumn{1}{c}{51} 
&
&\multicolumn{1}{c}{101}&\multicolumn{1}{c}{0}&\multicolumn{1}{c}{0}&\multicolumn{1}{c}{0}
\\
\multicolumn{1}{l}{povray\_r}&\multicolumn{1}{c}{4}&\multicolumn{1}{c}{0}&\multicolumn{1}{c}{0} &\multicolumn{1}{c}{0}&\multicolumn{1}{c}{4}&\multicolumn{1}{c}{0} 
&
&\multicolumn{1}{c}{0}&\multicolumn{1}{c}{0}&\multicolumn{1}{c}{4}&\multicolumn{1}{c}{0}
&
&\multicolumn{1}{c}{4}&\multicolumn{1}{c}{0}&\multicolumn{1}{c}{0}&\multicolumn{1}{c}{0} 
\\
\multicolumn{1}{l}{blender\_r}&\multicolumn{1}{c}{18}&\multicolumn{1}{c}{9}&\multicolumn{1}{c}{0}&\multicolumn{1}{c}{0}&\multicolumn{1}{c}{18} &\multicolumn{1}{c}{9}   
&
&\multicolumn{1}{c}{0}&\multicolumn{1}{c}{0}&\multicolumn{1}{c}{18} &\multicolumn{1}{c}{9}    
&
&\multicolumn{1}{c}{18}&\multicolumn{1}{c}{0}&\multicolumn{1}{c}{0}&\multicolumn{1}{c}{0}
\\
\cline{1-17}
\multicolumn{1}{l}{\textbf{Sum}}&\multicolumn{1}{c}{333}&\multicolumn{1}{c}{150}&\multicolumn{1}{c}{12} &\multicolumn{1}{c}{(12) 4/8}&\multicolumn{1}{c}{309}&\multicolumn{1}{c}{150}   
&%
&\multicolumn{1}{c}{9}&\multicolumn{1}{c}{(3) 2/8}&\multicolumn{1}{c}{321}&\multicolumn{1}{c}{150}  
&%
&\multicolumn{1}{c}{333}&\multicolumn{1}{c}{(0) 0/8}&\multicolumn{1}{c}{0} &\multicolumn{1}{c}{0}\\

\multicolumn{1}{l}{\textbf{Improvement}}&&&\colorbox{green}{96.4\%} &\multicolumn{1}{c}{50\%}&\multicolumn{1}{c}{92.8\%}&\colorbox{green}{100\%}   
&%
&\colorbox{green}{97.3\%}&\multicolumn{1}{c}{25\%}&\multicolumn{1}{c}{96.4\%}&\colorbox{green}{100\%}  
&%
&\multicolumn{1}{c}{}\\

\hline
\multicolumn{17}{l}{\textit{*\textbf{info included}: num of functions that have detailed context info in the standard library; \textbf{AC}: num of recovered return types which are the same as those from the standard libr-}}\\
\multicolumn{17}{l}{\textit{ary; \textbf{crash}: num of recovered return types which may cause the program to crash after static binary translation. In the \textbf{Sum} raw of \textbf{crash}, "()" refers to the total num of recov-}}\\
\multicolumn{17}{l}{\textit{ered return types that can lead to a crash. We also present the crash ratio of all tested benchmarks. \textbf{Improvement}: shows the improvement of EFACT compared to theirs.}}\\
\multicolumn{17}{l}{\textit{\textbf{EXF num} = \textbf{AC} + \textbf{crash} + \textbf{widening}.}}\\
\end{tabular}
\label{tab6}

}
\end{table*}
We delve deeper into the recovery of SPEC CPU® 2017 on \textit{HD\_x86}. First, we introduce the detailed performance of each part in EFACT. Table \ref{tab4} shows the distribution of VPC, FPC and MFC of each suite in SPEC CPU® 2017. In the ``this'' column of Table \ref{tab4}, 69.5\% mangled EXFs in SPEC CPU® 2017 will pass an implicit \textit{this} parameter, while all existing tools overlook to recover it. In the ``has ret'' column, only 13.5\% mangled EXFs can get a return type from the LLVM’s Demangling API.

Second, we perform an overall comparison between the correctness of the output from McSema, RetDec, and EFACT. Table \ref{tab5} shows the experiment result. Correctness is defined as follows:

\begin{Definition}
Correctness = (Number of completed EXFs that match the function declarations in the standard library) / (Number of EXFs in the target ELF file from \textit{glibc}, \textit{libstdc++} and \textit{libgfortran})
\label{definition2}
\end{Definition}
To be noticed, the EXF amount in Table \ref{tab5} is smaller than that in Table \ref{tab4}. This discrepancy is because, during our evaluation, only 34 of the total 43 benchmarks in SPEC CPU®2017 could be successfully analyzed by all three tools. Table \ref{tab5} reflects the evaluation results for these 34 benchmarks. In the ``variable parameter'' row, both RetDec and EFACT achieve 100\%. RetDec employs a recovery methodology similar to ours on recovering the variable-parameter function, but it does not consider the cross-ISA situation. McSema gets 0\%, indicating it does not address the VPC challenge; In the ``fixed parameter'' row, RetDec gets 89.7\%, performing well on EXFs from \textit{glibc} and \textit{libsdc++} but lacking coverage for the \textit{libfgortran} library. RetDec incorrectly recovers all the EXFs from \textit{libfgortran} library. McSema only gets 4.2\% on the FPC, treating all argument types as \textit{i64}, regardless of their actual types like a pointer or \textit{double}, which significantly reduces recovery accuracy. EFACT gets 100\% in this category; In the ``mangled'' row, both McSema and RetDec get a low percentage, with only 1.3\% and 4.7\% respectively. This category is a major contribution of our work. To further validate our tool's advancements in this respect, we have conducted a more detailed experiment in the next paragraph.
 
We evaluate the capabilities concerning the return-type recovery of mangled EXFs. In Table \ref{tab6}, we choose four dimensions to test the ability of mangled EXF recovery. ``AC'' represents whether the recovered return type is the same as those from the standard library; ``info loss'' means the recovered return type lacks context info, which is mentioned in Section \ref{MFC Solver}; ``crash'' indicates that the recovered return type may cause the program to crash after static binary translation. Crash is often triggered by casting a data type with more bits to a data type with less bits, which is mentioned in Section \ref{THREE CHALLANGE}. ``widening'' refers to scenarios when casting a data type with less bits to a data type with more bits. An example of this is when a function with a true return type of \textit{i32} is cast to a return type of \textit{i64}. In static binary translation, assigning an \textit{i32} to \textit{i64} doesn't induce data loss, but may affect the execution efficiency of the program after translation, as this might impact register allocation and memory layout. 

Table \ref{tab6} delineates a comparative analysis among RetDec, McSema, and EFACT on SPECrate 2017. From this table, we can see our tool performs best. A notable discrepancy is observed in the ``AC'' column for each of the tools, in which EFACT exhibits a remarkable improvement, accurately recovering 96.4\% more function return types than RetDec and 97.3\% more than McSema. This is mainly due to the simplistic method employed by RetDec and McSema in recovering the unknown return type with \textit{i32} and \textit{i64}, ignoring the fact that numerous mangled EXFs in SPECrate 2017 return a \textit{void}, \textit{double} or a pointer. RetDec and McSema have the same ratio of ``info loss'', as their recovery process does not concern the declaration context inherent to the source function declaration. The improvement of EFACT on ``info loss'' is 100\%. Besides, McSema experiences fewer crashes compared to RetDec, a consequence of the broader acceptance range of the \textit{i64} return type. RetDec may lose data when casting an \textit{i64} return type to \textit{i32}. Relatively, this policy also causes McSema to be more likely to trigger type widening. Based on our MFC Algorithm, our tool outperformed the other two in all aspects.

\subsection{Extensibility and usability}
\subsubsection{Extensibility}\label{ExtensibilityExp}
This paper mainly introduces how we implemented the lifting for EXFC on \textit{glibc}/\textit{libstdc++} in EFACT, which performs exceptionally well. In fact, our approach can be easily extended to support the lifting for binaries containing invocations to EXFs defined in other libraries, and binaries compiled from other programming languages.
 
To support an additional library, such as OpenSSL, we first identify the symbol that can be used to extract EXFs. In the ELF file using OpenSSL, symbols like ``\textit{@OPENSSL}'' are generated. We use this symbol to extract the EXF list. Second, we utilize the \textit{Dict} auto-generator introduced in Section \ref{librarydatabases} to construct a \textit{Dict} for OpenSSL. Finally, based on the EXF list and the newly generated \textit{Dict}, we follow the same workflow to complete the process. According to our evaluation, EFACT shows 100\% coverage on EXFs of the OpenSSL executable (which provides a command-line interface) in the OpenSSL library.
 
Although there may be significant differences between languages, the \textit{compile-link-and-run} chain is largely the same. The compiler of most languages generates an object file containing information akin to a symbol table. Our approach is effective as long as the ``symbol table'' is available. Our tool also supports recovering the \textit{libgfortran} library used in Fortran. We achieved 100\% coverage on the Fortran benchmarks in SPEC CPU® 2017. We also chose 10 command line utilities\footnote{bat, dust, fd, fend, hyperfine, miniserve, ripgrep, just, cargo-audit and cargo-wipe.} written in Rust. EFACT achieved a 100\% recovery rate for the EXFs from \textit{glibc}, which is used in the Rust utilities. 
\subsubsection{Usability}\label{UsabilityEXP}
As mentioned in Section \ref{introduciton}, to reduce the investment of binary lifting researchers in solving EXFC, we designed EFACT as a lightweight plugin. Our trials across various existing tools confirm that EFACT can adeptly aid them in resolving the EXFC problem. 

\textbf{Lasagne:} In Fig.~\ref{motivatingExample}, we can see Lasagne performs poorly in EXFC. To assess the improvement in Lasagne when combined with EFACT, we first use a binary compiled from the source code in Fig.~\ref{motivatingExample} as a test input. With EFACT's assistance, Lasagne successfully lifted the binary into LLVM IR. Second, we tested a handmade binary containing 20 EXFs, and Lasagne successfully lifted this binary as well. Third, for a more robust evaluation, we use the EEMBC benchmarks as test inputs. In this evaluation, EFACT's supplement helped Lasagne avoid the ``Unknown prototype for function'' error, but Lasagne failed in the subsequent instruction semantic translation step due to its lack of instruction coverage. Although Lasagne (combined with EFACT) could not lift the EEMBC benchmark completely, EFACT indeed helped Lasagne progress further in the binary lifting process.

\begin{figure}
\centerline{\includegraphics[width=\linewidth]{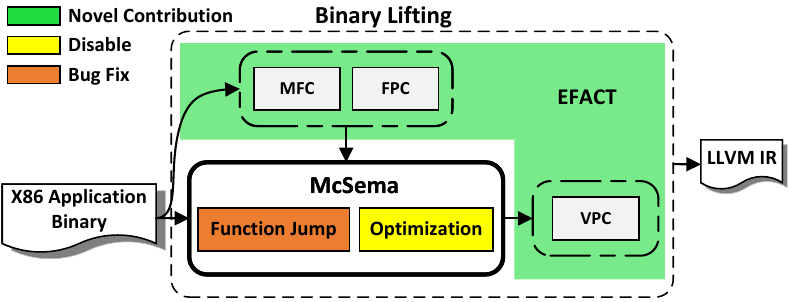}}
\caption{EFACT\_MC lifting workflow.}
\label{EFACTMCWORKFOLW}
\end{figure}

\textbf{McSema:} McSema is a well-known static binary lifting tool, which features an interface (\textit{--abi\_loader}) to accommodate function declarations in LLVM IR format. The EFACT output in LLVM IR format integrates well with this interface. To evaluate the improvements when combining McSema with EFACT, we developed a static binary translator, EFACT\_MC, which integrates McSema and EFACT. The framework of EFACT\_MC is shown in  Fig.~\ref{EFACTMCWORKFOLW}. We chose the EEMBC benchmarks to validate EFACT\_MC, using the original McSema for comparison. The specific details of this experiment are described in Section \ref{SBTEXP}.
\subsection{Static binary translation}\label{SBTEXP}
\begin{table}
\caption{EEMBC benchmark brief introduction.}

\renewcommand{\arraystretch}{1.2}
\resizebox{0.5\textwidth}{!}{
\begin{tabular}{lllll}

\hline
\multicolumn{1}{l}{\textbf{Benchmark}}&\multicolumn{1}{c}{\textbf{Introduction}} \\
\cline{1-2}
\multicolumn{1}{l}{CoreMark\_Pro 1.0} &\multicolumn{1}{l}{Sophisticated test of a processor's functionality} \\
\multicolumn{1}{l}{AutoBench 2.0}&\multicolumn{1}{l}{Multicore processor automotive workloads} \\
\multicolumn{1}{l}{MultiBench 1.1}&\multicolumn{1}{l}{Multicore processor integer workloads} \\
\multicolumn{1}{l}{FPMark 1.0}&\multicolumn{1}{l}{Multicore processor floating-point workloads}\\

\hline

\end{tabular}
\label{tab7}

}
\end{table}
Besides the above four dimensions, our tool has done a lot of work on solving the EXFC in the static binary translation area.
 
To evaluate our work, we employ an x86-64 to AArch64 (on Ubuntu 20.04) static binary translator, EFACT\_MC, which is based on McSema and EFACT. We use 4 benchmarks of EEMBC (build on \textit{HD\_x86} with Clang11) as input binaries. Table \ref{tab7} provides a brief introduction to them (other benchmarks require mobile hardware for installation or are not buildable due to insufficient documentation support). Our evaluation method consists of the following steps: First, we build the EEMBC benchmark and obtain the test binary (on \textit{HD\_x86}). Second, we put the binary into both McSema and EFACT\_MC, obtaining the lifted LLVM IR (on \textit{HD\_x86}). Then, we use Clang11 to compile the lifted LLVM IR to ELF (on \textit{HD\_ARM} and \textit{HD\_x86}). Finally, we test the correctness of recompiled ELF from McSema and EFACT\_MC, respectively. All EEMBC benchmarks output their execution results to the terminal. We compare these results (outputs from the EEMBC benchmarks compiled from source code) with those from EFACT\_MC and McSema to verify the correctness of the translated binary.
 
Fig.~\ref{EFACTMCWORKFOLW} shows an overview of the binary lifting process facilitated by EFACT\_MC. We use EFACT as a plugin to help McSema better solve the EXFC problem. To minimize the potential impact from other factors, we fix several bugs in the function jump part of McSema, and disable all the default optimization options to avoid potential interfering factors. We also apply the above modifications to the McSema which is used as a comparison baseline.
\begin{table}
\caption{EFACT\_MC's evaluation result (x86-64 to x86-64).}

\renewcommand{\arraystretch}{1.2}
\resizebox{0.5\textwidth}{!}{
\begin{tabular}{lcccccccccc}

\hline
 \multirow{2}*{\textbf{Benchmark}}& \multirow{2}*{\textbf{Test case}}&\multicolumn{4}{c}{\textbf{McSema}}&\multicolumn{4}{c}{\textbf{EFACT\_MC}}&\multirow{2}*{\textbf{Improvement}} \\
\cline{3-6}
\cline{7-10}
&\multicolumn{1}{c}{\textbf{amount}}&\multicolumn{1}{c}{\textbf{Pass}}&\multicolumn{1}{c}{\textbf{\circled{1}}}&\multicolumn{1}{c}{\textbf{\circled{2}}}&\multicolumn{1}{c}{\textbf{\circled{3}}}&\multicolumn{1}{c}{\textbf{Pass}}&\multicolumn{1}{c}{\textbf{\circled{1}}}&\multicolumn{1}{c}{\textbf{\circled{2}}}&\multicolumn{1}{c}{\textbf{\circled{3}}} \\
\cline{1-11}
\multicolumn{1}{l}{CoreMark\_Pro}&\multicolumn{1}{c}{9}&\multicolumn{1}{c}{6}&\multicolumn{1}{c}{1}&\multicolumn{1}{c}{0}&\multicolumn{1}{c}{2}&\multicolumn{1}{c}{8}&\multicolumn{1}{c}{1}&\multicolumn{1}{c}{0}&\multicolumn{1}{c}{0}&\multicolumn{1}{c}{22.2\%} \\
\multicolumn{1}{l}{AutoBench}&\multicolumn{1}{c}{20}&\multicolumn{1}{c}{12}&\multicolumn{1}{c}{0}&\colorbox{red}{2}&\multicolumn{1}{c}{6}&\multicolumn{1}{c}{20}&\multicolumn{1}{c}{0}&\colorbox{green}{0}&\multicolumn{1}{c}{0}&\colorbox{green}{40.0\%} \\
\multicolumn{1}{l}{MultiBench}&\multicolumn{1}{c}{28}&\multicolumn{1}{c}{23}&\multicolumn{1}{c}{0}&\multicolumn{1}{c}{0}&\multicolumn{1}{c}{5}&\multicolumn{1}{c}{28}&\multicolumn{1}{c}{0}&\multicolumn{1}{c}{0}&\multicolumn{1}{c}{0}&\multicolumn{1}{c}{17.9\%} \\
\multicolumn{1}{l}{FPmark}&\multicolumn{1}{c}{52}&\colorbox{red}{21}&\multicolumn{1}{c}{6}&\multicolumn{1}{c}{0}&\multicolumn{1}{c}{25}&\colorbox{green}{46}&\multicolumn{1}{c}{6}&\multicolumn{1}{c}{0}&\multicolumn{1}{c}{0}&\colorbox{green}{48.1\%} \\
\cline{1-11}
\multicolumn{1}{l}{\textbf{Sum}}&\multicolumn{1}{c}{109}&\multicolumn{1}{c}{62}&\multicolumn{1}{c}{7}&\multicolumn{1}{c}{2}&\multicolumn{1}{c}{38}&\multicolumn{1}{c}{102}&\multicolumn{1}{c}{7}&\multicolumn{1}{c}{0}&\multicolumn{1}{c}{0}&\colorbox{green}{36.7\%} \\

\hline
\multicolumn{11}{l}{\textit{*\textbf{Pass}: successfully run without bugs. \textbf{\circled{1}}: fail to be recompiled and linked to an ELF. \textbf{\circled{2}}: suc-}}\\
\multicolumn{11}{l}{\textit{cessfully generate the ELF, but some bugs crash the program while running. \textbf{\circled{3}}: successfully}}\\
\multicolumn{11}{l}{\textit{generate the ELF, but have some bugs that do not crash the program while running.}}\\

\end{tabular}
\label{tab8}

}
\end{table}

\begin{table}
\caption{EFACT\_MC's evaluation result (x86-64 to AArch64).}

\renewcommand{\arraystretch}{1.2}
\resizebox{0.5\textwidth}{!}{
\begin{tabular}{lcccccccccc}

\hline
 \multirow{2}*{\textbf{Benchmark}}& \multirow{2}*{\textbf{Test case}}&\multicolumn{4}{c}{\textbf{McSema}}&\multicolumn{4}{c}{\textbf{EFACT\_MC}}&\multirow{2}*{\textbf{Improvement}} \\
\cline{3-6}
\cline{7-10}
&\multicolumn{1}{c}{\textbf{amount}}&\multicolumn{1}{c}{\textbf{Pass}}&\multicolumn{1}{c}{\textbf{\circled{1}}}&\multicolumn{1}{c}{\textbf{\circled{2}}}&\multicolumn{1}{c}{\textbf{\circled{3}}}&\multicolumn{1}{c}{\textbf{Pass}}&\multicolumn{1}{c}{\textbf{\circled{1}}}&\multicolumn{1}{c}{\textbf{\circled{2}}}&\multicolumn{1}{c}{\textbf{\circled{3}}} \\
\cline{1-11}
\multicolumn{1}{l}{CoreMark\_Pro}&\multicolumn{1}{c}{9}&\multicolumn{1}{c}{0}&\multicolumn{1}{c}{1}&\multicolumn{1}{c}{8}&\multicolumn{1}{c}{0}&\multicolumn{1}{c}{8}&\multicolumn{1}{c}{1}&\multicolumn{1}{c}{0}&\multicolumn{1}{c}{0}&\multicolumn{1}{c}{88.9\%} \\
\multicolumn{1}{l}{AutoBench}&\multicolumn{1}{c}{20}&\multicolumn{1}{c}{0}&\multicolumn{1}{c}{0}&\multicolumn{1}{c}{20}&\multicolumn{1}{c}{0}&\multicolumn{1}{c}{20}&\multicolumn{1}{c}{0}&\multicolumn{1}{c}{0}&\multicolumn{1}{c}{0}&\multicolumn{1}{c}{100\%} \\
\multicolumn{1}{l}{MultiBench}&\multicolumn{1}{c}{28}&\multicolumn{1}{c}{0}&\multicolumn{1}{c}{0}&\multicolumn{1}{c}{28}&\multicolumn{1}{c}{0}&\multicolumn{1}{c}{28}&\multicolumn{1}{c}{0}&\multicolumn{1}{c}{0}&\multicolumn{1}{c}{0}&\multicolumn{1}{c}{100\%} \\
\multicolumn{1}{l}{FPmark}&\multicolumn{1}{c}{52}&\multicolumn{1}{c}{0}&\multicolumn{1}{c}{6}&\multicolumn{1}{c}{46}&\multicolumn{1}{c}{0}&\multicolumn{1}{c}{46}&\multicolumn{1}{c}{6}&\multicolumn{1}{c}{0}&\multicolumn{1}{c}{0}&\multicolumn{1}{c}{88.5\%} \\
\cline{1-11}
\multicolumn{1}{l}{\textbf{Sum}}&\multicolumn{1}{c}{109}&\colorbox{red}{0}&\multicolumn{1}{c}{7}&\multicolumn{1}{c}{102}&\multicolumn{1}{c}{0}&\colorbox{green}{102}&\multicolumn{1}{c}{7}&\multicolumn{1}{c}{0}&\multicolumn{1}{c}{0}&\colorbox{green}{93.6\%} \\

\hline
\multicolumn{11}{l}{\textit{*\textbf{Pass}: successfully run without bugs. \textbf{\circled{1}}: fail to be recompiled and linked to an ELF.\textbf{ \circled{2}}: suc-}}\\
\multicolumn{11}{l}{\textit{cessfully generate the ELF, but some bugs crash the program while running. \textbf{\circled{3}}: successfully}}\\
\multicolumn{11}{l}{\textit{generate the ELF, but have some bugs that do not crash the program while running.}}\\
\end{tabular}
\label{tab9}

}
\end{table}
Table \ref{tab8} and Table \ref{tab9} show the evaluation result. We categorize four evaluation indexes: Pass, \circled{1}, \circled{2}, and \circled{3}. \circled{1} mainly results from McSema's ability in control-flow-graph recovery or x86-64 instruction coverage, which is not related to the EXFC problem. \circled{2} and \circled{3} are strongly related to our focus because incorrect parameter passing to the dynamically linked shared library or a wrong return type from the shared library can both cause the translated program to crash or encounter errors.
 
We first compare the results that choose x86-64 as the target ISA. The goal of this series of experiments is to sidestep the differences between ISAs, allowing us to more accurately gauge the enhancements that complete function declaration recovery can introduce to static binary translation. From Table \ref{tab8}, we can see EFACT\_MC successfully translates 36.7\% more benchmarks than McSema (without EFACT). Particularly, within the confines of the green rectangle in AutoBench, our tool has been instrumental in solving program crashes, proving our earlier conclusion that the EXFC problem can indeed lead to a program crash. For FPmark, there is a significant improvement in the translation rate. FPmark is a floating-point workload, it contains many functions that use \textit{double} as return type and parameters. The default EXFC policy employed by McSema—completing \textit{double} with \textit{int}—triggers serious problems.

Then we compare the results that choose AArch64 as the target ISA, where we intend to highlight that the differences between ISAs necessitate considerable modifications, as showcased in this series of tests. From Table \ref{tab9}, we can see McSema failed to translate all the benchmarks, This is because the source code of EEMBC uses a lot of \textit{va\_list} to display messages. However, with the help of our LLVM Passes mentioned in Section \ref{VPC SOLVER}, EFACT\_MC successfully translates 93.6\% more benchmarks than McSema. 

From the column \circled{1} in Table \ref{tab8} and Table \ref{tab9}, EFACT\_MC shows no improvement. All seven of these failed cases report errors due to encountering unsupported x86-64 instructions while lifting the source ELF. These errors are unrelated to EXFC. In future developments, we aim to improve EFACT\_MC's coverage on x86-64 instructions.
 
Overall, EFACT greatly improves the accuracy of static binary translation.

\subsection{EXF's perspective on SPEC CPU® 2017}\label{EXFEXP}
To help binary lifting researchers better understand the EXFC problem, we have analyzed all the EXFs in SPEC CPU® 2017, without library restrictions, and we chose ISA, optimization option, and processor architecture (three frequently concerned aspects in binary lifting) for characterization. It should be noted that the counts of EXFs (on \textit{HD\_x86}, 64-bit) in Table \ref{tab10}, Table \ref{tab11}, and  Table \ref{tab12} exceed those in Table \ref{tab4}. While Table \ref{tab4} displays EXFs souring from \textit{glibc}, \textit{libstdc++} and \textit{libgfortran}, the latter three tables count all EXFs in the benchmark. This includes libraries beyond the aforementioned three. For instance, \textit{620.omnetpp\_s} contain EXFs sourced from ``\textit{@OMP}''. 

\begin{table}
\centering
\caption{Total EXF counts of SPEC CPU® 2017 on different ISA (\textit{64-bit} ,\textit{O0}).}

\renewcommand{\arraystretch}{1.2}
\resizebox{0.41\textwidth}{!}{
\begin{tabular}{lcccc}

\hline
\multicolumn{1}{c}{\textbf{Suite}}&\multicolumn{1}{c}{\textbf{HD\_x86}}&&\multicolumn{1}{c}{\textbf{HD\_ARM}}&\multicolumn{1}{c}{\textbf{difference}}\\
\hline
intspeed&\multicolumn{1}{c}{931}&&\multicolumn{1}{c}{939}&\multicolumn{1}{c}{0.9\%}\\
intrate&\multicolumn{1}{c}{923}&&\multicolumn{1}{c}{934}&\multicolumn{1}{c}{1.2\%}\\
fpspeed&\multicolumn{1}{c}{1004}&&\multicolumn{1}{c}{1016}&\multicolumn{1}{c}{1.2\%}\\
fprate&\multicolumn{1}{c}{1374}&&\multicolumn{1}{c}{1407}&\multicolumn{1}{c}{2.4\%}\\
\cline{1-5}
\textbf{Sum}&\multicolumn{1}{c}{4232}&&\multicolumn{1}{c}{4296}&\multicolumn{1}{c}{1.5\%}\\

\hline

\multicolumn{5}{l}{\textit{ *\textbf{difference}: how many EXFs HD\_ARM have than HD\_x86.}}\\
\end{tabular}
\label{tab10}

}
\end{table}
 
First of all, from Table \ref{tab10}, we can see there is a significant amount of EXFs that cannot be ignored. The number of EXFs in \textit{intspeed} is nearly the same as \textit{intrate} because these two suites have the same benchmark numbers and application areas, but differ in their performance calculation methods. The small difference mainly lies in their different output format (e.g. \textit{505.mcf\_r} use more output functions than \textit{605.mcf\_s}, like \textit{fprintf}) and their measurement methods (e.g. \textit{657.xz\_s} use more functions to get measurement matrix than \textit{557.xz\_r}, like \textit{omp\_get\_max\_threads}). \textit{fpspeed} and \textit{fprate} differ so much because \textit{fprate} has 3 more benchmarks than \textit{fpspeed}.
 
Second, we observe that different ISAs impact the generated EXF numbers. \textit{HD\_ARM} has 1.5\% more EXFs than \textit{HD\_x86}. Upon further investigation, we found that \textit{HD\_ARM} requires an additional \textit{abort} function in each generated ELF compared to those on \textit{HD\_x86}. This is because AArch64 employs an \textit{abort} function to handle scenarios where the main program (or \textit{\_\_libc\_start\_main}) returns unexpectedly. In contrast, x86-64 uses the \textit{HLT} instruction for this purpose. Furthermore, \textit{HD\_ARM} requires more functions to execute the same floating-point operations than x86-64 (owing to AArch64's RISC architecture).
\begin{table}
\centering
\caption{Total EXF counts of SPEC CPU® 2017 with different optimization option (\textit{HD\_x86}, \textit{64-bit}).}

\renewcommand{\arraystretch}{1.2}
\resizebox{0.33\textwidth}{!}{
\begin{tabular}{lcccc}

\hline
\multicolumn{1}{c}{\textbf{Suite}}&\multicolumn{1}{c}{\textbf{O0}}&&\multicolumn{1}{c}{\textbf{O3}}&\multicolumn{1}{c}{\textbf{difference}}\\
\hline
intspeed&\multicolumn{1}{c}{931}&&\multicolumn{1}{c}{859}&\multicolumn{1}{c}{-7.7\%}\\
intrate&\multicolumn{1}{c}{923}&&\multicolumn{1}{c}{853}&\multicolumn{1}{c}{-7.6\%}\\
fpspeed&\multicolumn{1}{c}{1004}&&\multicolumn{1}{c}{1043}&\multicolumn{1}{c}{3.9\%}\\
fprate&\multicolumn{1}{c}{1374}&&\multicolumn{1}{c}{1314}&\multicolumn{1}{c}{-4.4\%}\\
\cline{1-5}
\textbf{Sum}&\multicolumn{1}{c}{4232}&&\multicolumn{1}{c}{4069}&\multicolumn{1}{c}{-3.9\%}\\

\hline

\multicolumn{5}{l}{\textit{*\textbf{difference}: how many EXFs O3 have than O0.}}\\
\end{tabular}
\label{tab11}

}
\end{table}
 
Third, from Table \ref{tab11}, we can see the optimization option also impacts the number of generated EXFs. For example, from \textit{intspeed}, \textit{intrate} and \textit{fprate} of Table \ref{tab11}, we can see sometimes a higher optimization option decreases the number of generated EXFs. For example, when compiling \textit{657.xz\_s} with the \textit{O3} optimization option, it will result in the inlining of the \textit{strlen} function, thereby reducing the number of EXFs. However, it's not always accurate to state that a higher compiler optimization option leads to fewer EXFs. Table \ref{tab11} shows an increase in EXFs under the \textit{O3} optimization option compared to \textit{O0} for \textit{fpspeed}. Further analysis shows that compiling \textit{621.wrf\_s} with the \textit{O3} optimization option causes more EXFs to be generated, including a group of functions like \textit{\_ZGVdN4v\_sin} from \textit{Libmvec}. \textit{Libmvec} is an x86-64 \textit{glibc} library incorporating SSE4, AVX, AVX2, and AVX-512 vectorized functions for mathematical operations like \textit{cos}, \textit{exp}, \textit{sin}, etc. On the other hand, when \textit{621.wrf\_s} is compiled with the \textit{O0} optimization option, the compiler will not enable the Vectorization optimization, therefore, the generated binary will not include EXFs from \textit{Libmvec}, leading to a reduced number of EXFs.

\begin{table}
\centering
\caption{Total EXF counts of SPEC CPU® 2017 on different process architecture (\textit{HD\_x86} ,\textit{O0}).}

\renewcommand{\arraystretch}{1.2}
\resizebox{0.33\textwidth}{!}{
\begin{tabular}{lcccc}

\hline
\multicolumn{1}{c}{\textbf{Suite}}&\multicolumn{1}{c}{\textbf{32-bit}}&&\multicolumn{1}{c}{\textbf{64-bit}}&\multicolumn{1}{c}{\textbf{difference}}\\
\hline
intspeed&\multicolumn{1}{c}{-}&&\multicolumn{1}{c}{931}&\multicolumn{1}{c}{-}\\
intrate&\multicolumn{1}{c}{923}&&\multicolumn{1}{c}{923}&\multicolumn{1}{c}{0\%}\\
fpspeed&\multicolumn{1}{c}{-}&&\multicolumn{1}{c}{1004}&\multicolumn{1}{c}{-}\\
fprate&\multicolumn{1}{c}{1385}&&\multicolumn{1}{c}{1374}&\multicolumn{1}{c}{-0.8\%}\\
\cline{1-5}
\textbf{Sum}&\multicolumn{1}{c}{2308}&&\multicolumn{1}{c}{2297}&\multicolumn{1}{c}{-0.5\%}\\

\hline
\multicolumn{5}{l}{\textit{*raw \textbf{Sum} calculate intrate and fprate. \textbf{differe-}}}\\
\multicolumn{5}{l}{\textit{\textbf{nce:} how many EXFs 64-bit have than 32-bit.}}\\
\end{tabular}
\label{tab12}

}
\end{table}
 
Fourth, Table \ref{tab12} shows the EXF numbers for both 32-bit and 64-bit benchmarks. Only the SPECrate suites support \textit{-m32} to generate 32-bit ELF files\footnote{https://www.spec.org/cpu2017/Docs/overview.html}. We conduct this experiment on \textit{HD\_x86}. From the table, we can see the 32-bit benchmarks have 0.5\% more EXFs than the 64-bit benchmarks. When building the 64-bit benchmarks, the compiler usually links with newer library versions than the 32-bit ones. Newer version libraries have some default optimizations that may inline parts of the internal implementation of certain functions (transparent to the programmer) to make the program faster, resulting in fewer EXFs.
 
In summary, our analysis confirms that EXFC is an inevitable challenge. Factors such as ISA, optimization option, and processor architecture directly impact the number of generated EXFs. Our study reveals that in SPEC CPU\circledR 2017, AArch64 binaries have 1.5\% more EXFs than x86-64 binaries for SPEC CPU® 2017, due to its weak memory model and RISC architecture; 32-bit benchmarks have 0.5\% more EXFs than 64-bit benchmarks as 64-bit benchmarks link to newer version libraries, which likely inline some implementation functions, resulting in fewer EXFs. We recommend that binary lifting researchers focusing on the ARM platform and 32-bit processor architecture should pay more attention to EXFC.

\section{Limitations}

The current version of EFACT has several limitations:
 
\textbf{(1)} EFACT cannot support the shared library if the library's source files, in which the functions are declared, are missing.
 
\textbf{(2)} Extra manual effort is needed for a few libraries if their source files, in which the functions are declared, are not well organized, like \textit{libgfortran}. We have automated much of this process through scripts.
 
\textbf{(3)} Although EFACT demonstrates 100\% coverage in the experiments described in Section \ref{CoverageExp}, it's important to note that our focus has been on covering benchmarks like EEMBC and SPEC CPU\circledR 2017. While achieving full coverage on these benchmarks suggests practical usability, there may still exist functions, in GNU standard libraries, that EFACT fails to recover. EFACT provides a manual insertion interface, allowing users to add EXFs that the current version may not support. With more practical workloads explored by us using EFACT, its coverage on EXFC will be continuously improved accordingly. 
 
\textbf{(4)} Our current LLVM Passes have only been verified for compatibility with the LLVM IR lifted by McSema. Our VPC solver is not guaranteed to recover all the variable-parameter functions. Supporting the EXPC for more functions with variable parameters is one of our future plans.

\section{Related work}

In this section, we first discuss function signature recovery. Then, we discuss dynamic binary lifting. Finally, we shift our focus to the related works of EFACT and divide them into three main categories based on output format: binary-to-ASM, binary-to-IR, and binary-to-high-level languages.
\subsection{Function signature recovery}
Function signature recovery~\cite{7546543,5288c9a1d7c54ac9a168245dbc65a90b,DBLP:conf/sp/LinG21}~aims to recover the number and types of arguments of a function from binary executables, where ``arguments'' refer to the function's parameters, which does not include the return type. Different from EFACT's application domain, function signature recovery is used to construct a fine-grained control-flow graph (CFG) for aiding control-flow integrity (CFI) enforcement. It operates at the binary-to-ASM level and is used for defending against control-flow hijacking attacks, ensuring that control flows only occur between callees and callers with matching function signatures.
\subsection{Dynamic binary lifting}
Before 2020, binary lifting primarily appeared in static frameworks. Although many dynamic frameworks at that time~\cite{DBLP:conf/csse/GuanLLL08,hong2012hqemu,rokicki2018hybrid} had processes for translating binaries into IR, researchers did not specifically propose the concept of dynamic binary lifting due to the presence of interpreters in dynamic frameworks. In 2020, BinRec~\cite{DBLP:conf/eurosys/AltinayNKRZDGNV20} introduced the concept of Dynamic Binary Lifting, presenting a detailed analysis of how dynamic environments could address the issues present in previous static binary lifting methods.
\subsection{Static binary lifting tools}

\begin{table}
\centering
\caption{Related work overview (static phase).}

\renewcommand{\arraystretch}{1.2}
\resizebox{0.38\textwidth}{!}{
\begin{tabular}{lcccc}

\hline
 \multirow{2}*{\textbf{Tool}}&\multirow{2}*{\textbf{Output}}&\multicolumn{3}{c}{\textbf{EXFC}} \\
\cline{3-5}
&&\multicolumn{1}{c}{\textbf{FPC}}&\multicolumn{1}{c}{\textbf{VPC}}&\multicolumn{1}{c}{\textbf{MFC}} \\

\hline
EFACT&\multicolumn{1}{c}{C/C++,LLVM IR}&\multicolumn{1}{c}{$\color{green}\checkmark$}&\multicolumn{1}{c}{$\color{green}\checkmark$}&\multicolumn{1}{c}{$\color{green}\checkmark$}\\
McSema&\multicolumn{1}{c}{LLVM IR}&\multicolumn{1}{c}{$\color{blue}\checkmark\kern-1.1ex\raisebox{.7ex}{\rotatebox[origin=c]{125}{--}}$}&\multicolumn{1}{c}{$\color{blue}\checkmark\kern-1.1ex\raisebox{.7ex}{\rotatebox[origin=c]{125}{--}}$}&\multicolumn{1}{c}{$\color{blue}\checkmark\kern-1.1ex\raisebox{.7ex}{\rotatebox[origin=c]{125}{--}}$}\\
SSM&\multicolumn{1}{c}{ASM}&\multicolumn{1}{c}{$\color{red}\scalebox{0.8}{\usym{2613}}$}&\multicolumn{1}{c}{$\color{red}\scalebox{0.8}{\usym{2613}}$}&\multicolumn{1}{c}{$\color{red}\scalebox{0.8}{\usym{2613}}$}\\
SmartDec&\multicolumn{1}{c}{C++}&\multicolumn{1}{c}{$\color{red}\scalebox{0.8}{\usym{2613}}$}&\multicolumn{1}{c}{$\color{red}\scalebox{0.8}{\usym{2613}}$}&\multicolumn{1}{c}{$\color{red}\scalebox{0.8}{\usym{2613}}$}\\
Ghidra&\multicolumn{1}{c}{C/C++,ASM}&\multicolumn{1}{c}{$\color{green}\checkmark$}&\multicolumn{1}{c}{$\color{blue}\checkmark\kern-1.1ex\raisebox{.7ex}{\rotatebox[origin=c]{125}{--}}$}&\multicolumn{1}{c}{$\color{blue}\checkmark\kern-1.1ex\raisebox{.7ex}{\rotatebox[origin=c]{125}{--}}$}\\
RetDec&\multicolumn{1}{c}{C,LLVM IR}&\multicolumn{1}{c}{$\color{green}\checkmark$}&\multicolumn{1}{c}{$\color{blue}\checkmark\kern-1.1ex\raisebox{.7ex}{\rotatebox[origin=c]{125}{--}}$}&\multicolumn{1}{c}{$\color{blue}\checkmark\kern-1.1ex\raisebox{.7ex}{\rotatebox[origin=c]{125}{--}}$}\\
Lasagne&\multicolumn{1}{c}{LLVM IR}&\multicolumn{1}{c}{$\color{red}\scalebox{0.8}{\usym{2613}}$}&\multicolumn{1}{c}{$\color{blue}\checkmark\kern-1.1ex\raisebox{.7ex}{\rotatebox[origin=c]{125}{--}}$}&\multicolumn{1}{c}{$\color{red}\scalebox{0.8}{\usym{2613}}$}\\
IDA PRO&\multicolumn{1}{c}{C/C++,ASM}&\multicolumn{1}{c}{$\color{green}\checkmark$}&\multicolumn{1}{c}{$\color{blue}\checkmark\kern-1.1ex\raisebox{.7ex}{\rotatebox[origin=c]{125}{--}}$}&\multicolumn{1}{c}{$\color{blue}\checkmark\kern-1.1ex\raisebox{.7ex}{\rotatebox[origin=c]{125}{--}}$}\\

\hline

\multicolumn{5}{l}{\textit{*$\color{green}\checkmark$: correctly recover EXF and match all the require-}}\\
\multicolumn{5}{l}{\textit{ments of the challenge. $\color{blue}\checkmark\kern-1.1ex\raisebox{.7ex}{\rotatebox[origin=c]{125}{--}}$: can generate a recovered}}\\
\multicolumn{5}{l}{\textit{result, but failed to match all the response challenge'}}\\
\multicolumn{5}{l}{ requirements. \textit{$\color{red}\scalebox{0.8}{\usym{2613}}$ :can not generate a recovered result.}}\\

\end{tabular}
\label{tab13}

}
\end{table}

Table \ref{tab13} presents the summary of existing tools’ capabilities in addressing the EXFC problem. We will present an overview on each of them.
\subsubsection{\textbf{Binary-to-ASM}}
ASM can accurately represent the behavior of the source binary. However, ASM lacks direct information on function parameters and return type, making it difficult to supplement the details of function calls. IDA PRO~\cite{hex2014ida} and Ghidra~\cite{Ghidra} are two well-established tools that not only output the source binary's ASM but also provide a transformation interpretation in C/C++ program format. There are still shortcomings when applied to the EXFC issue, especially the return type of mangled EXF.
\subsubsection{\textbf{Binary-to-IR}}
Many static binary frameworks utilize LLVM IR or TCG IR~\cite{bellard2005qemu} for translation or analysis, which is highly relevant to our work and has significant potential benefits. After trying to obtain code links and contacting corresponding authors of relevant papers via email, both LLBT~\cite{shen2012llbt} and SecondWrite~\cite{anand2013compiler} were unable to reproduce their work. Moreover, neither of their papers mentioned the EXFC issue. McSema and RetDec, as mentioned in Section \ref{introduciton}, they have respective shortcomings. Lasagne experienced significant problems with FPC and MFC. Despite adding \textit{cmath.h} to the ``\textit{--include-files}'' option, Lasagne is still unable to lift \textit{cos}. Incorrect supplementation of function parameters during static binary translation can result in parameter loss or type errors, potentially leading to a program crash when translating complex software. Existing tools all failed to recover the implicit \textit{this} parameter in MFC. Besides our tool, RetDec performs best in this area. However, it cannot provide the accurate return type for mangled EXF.
\subsubsection{\textbf{Binary-to-high-level languages}}
Few tools are capable of producing quality high-level languages such as C and C++, as doing so requires a significant amount of information and engineering effort. IDA PRO and Ghidra generate the most accurate output in C format. Phoenix~\cite{schwartz2013native}, FoxDec~\cite{verbeek2020sound}, and SmartDec~\cite{fokin2011smartdec} (C++ format) are three other published lifting tools. Unfortunately, these tools haven't discussed the EXFC problem in their papers. Additionally, Phoenix is not an open-source tool, and FoxDec's module responsible for generating C language output remains closed-source. SmartDec has shown poor performance on EXFC.

\section{Conclusion}
In this paper, we present EFACT, an External Function Auto-Completion Tool for static binary lifting frameworks. EFACT can recover the full function declarations of an ELF'EXFs, unbounded by ISA, processor architecture, OS, and library version. Notably, our tool can recover the implicit \textit{this} parameter and return type of mangled EXF with the help of our MFC algorithm. EFACT automatically generates dictionaries (\textit{Dicts}) to cover popularly used external function declarations.
 
EFACT demonstrates 100\% coverage of EXFs derived from the \textit{glibc}, \textit{libstdc++} and \textit{libgfortran} on SPEC CPU® 2017. Notably, in the context of MFC, EFACT exhibits a remarkable improvement, accurately recovering 96.4\% more function return type than RetDec and 97.3\% more than McSema on SPECrate 2017. EFACT generates output in C/C++ program format and LLVM IR format to better assist other binary rewriting frameworks. It has been tested and proven compatible with Lasagne and McSema.

Additionally, we delve further into the static binary translation domain, addressing several cross-ISA EXFC challenges. We built a static binary translator called EFACT\_MC, which integrates EFACT into McSema and translates binary from x86-64 to AArch64 (on Ubuntu 20.04). EFACT\_MC's performance evaluation showed a notable increase in translation accuracy, with 36.7\% more correct translations in EEMBC benchmarks when moving from x86-64 to x86-64, and 93.6\% more from x86-64 to AArch64, compared to the original McSema.

EFACT is designed to be extensible, supporting not only binaries compiled from C/C++ but also from languages like Fortran and Rust. Additionally, it can be adapted to work with libraries like OpenSSL.
 
We characterized the SPEC CPU® 2017 benchmark from the perspective of EXFs, demonstrating that EXFC is an inevitable challenge. We found that ISA, optimization option, and processor architecture directly impact the number of generated EXFs. Based on our analytical results, we recommend that binary lifting researchers focusing on the ARM platform and 32-bit processor architecture should pay more attention to EXFC.
 
We believe our work on EFACT helps push forward static binary lifting research and we plan to continue enhancing the EXFC ability of EFACT. Currently, EFACT\_MC primarily serves to evaluate the EXFC accuracy of our framework. In our experiments on cross-platform static binary translation, we observe that the detailed implementations of various data types (such as \textit{long}) and data structures (such as \textit{va\_list}, as mentioned in Section \ref{VPC SOLVER}) vary across platforms. These differences can lead to the crash of the translated program when being executed. We refer to these issues as the cross-platform data interpretation problem. To address this problem, we are in the process of developing a specific LLVM Pass. This Pass will perform data morphing/synchronization before/after certain external function calls within the lifted LLVM IR. Our goal is to ensure that the data types and structures are compatible with the target platform once the lifted LLVM IR is recompiled into an ELF file, thereby solving the problem as much as possible.

\subsection*{Acknowledgments}
This work is supported by the National Natural Science Foundation of China (No. 62272167).
\subsection*{Data availability}
The source code of EFACT is available at \url{https://github.com/solecnugit/EFACT}.


\bibliographystyle{model3-num-names}

\bibliography{cas-refs}
\quad\\
\footnotesize {\textbf{Yilei Zhang} received his bachelor's degree from Nanjing Agricultural University in Jiangsu, China, in 2019. He is currently pursuing his PhD at the School of Data Science and Engineering, East China Normal University. His research interests include binary translation.}
\quad\\
\footnotesize {\textbf{Haoyu Liao}
received his bachelor's degree from Chong Qing Jiao Tong University in 2022. He is pursuing a graduate degree from the School of Data Science and Engineering, East China Normal University. His research interests include performance evaluation and analysis.}
\quad\\
\footnotesize {\textbf{Zekun Wang}
received his bachelor's degree from Northeastern University in 2022. He is pursuing a graduate degree from the School of Data Science and Engineering, East China Normal University, Shanghai, China. His research interests include workload sampling and analysis.}
\quad\\
\footnotesize {\textbf{Bo Huang}
received his PhD degree in computer science from Fudan University, China, in 2000. He is a professor with East China Normal University. With 20+ years' industry experience, his current research interests include compiler technology and data-driven system optimization.}
\quad\\
\footnotesize {\textbf{Jianmei Guo}
is now a Professor at East China Normal University. His research interests include the quality assurance and performance optimization of software systems. He received his Ph.D. in Computer Science in 2011 from Shanghai Jiao Tong University. He was a Postdoctoral Fellow at the University of Waterloo, an Associate Professor at East China University of Science and Technology, and a Staff Engineer at Alibaba Group. He received two Best Paper Awards of SPLC 2016, Canadian AI 2017, and an ACM SIGSOFT Distinguished Paper Award of ASE 2015. He is a member of ACM, IEEE and CCF. }

\end{document}